\documentclass[11pt]{article}

\usepackage{fullpage}
\usepackage{setspace}
\usepackage{parskip}
\usepackage{titlesec}
\usepackage{xcolor}
\usepackage{lineno}
\usepackage{chngcntr}

\usepackage{color}

\PassOptionsToPackage{hyphens}{url}
\usepackage[colorlinks = true,
            linkcolor = blue,
            urlcolor  = blue,
            citecolor = blue,
            anchorcolor = blue]{hyperref}
\usepackage{etoolbox}
\makeatletter
\patchcmd\@combinedblfloats{\box\@outputbox}{\unvbox\@outputbox}{}{%
  \errmessage{\noexpand\@combinedblfloats could not be patched}%
}%
\makeatother

\renewenvironment{abstract}
  {{\bfseries\noindent{\abstractname}\par\nobreak}\footnotesize}
  {\bigskip}

\titlespacing{\section}{0pt}{*3}{*1}
\titlespacing{\subsection}{0pt}{*2}{*0.5}
\titlespacing{\subsubsection}{0pt}{*1.5}{0pt}

\usepackage{authblk}

\usepackage[super]{natbib}
\setcitestyle{citesep={,}}
\usepackage{graphicx}
\usepackage[space]{grffile}
\usepackage{latexsym}
\usepackage{textcomp}
\usepackage{longtable}
\usepackage{tabulary}
\usepackage{booktabs,array,multirow}
\usepackage{amsfonts,amsmath,amssymb}
\providecommand\citet{\cite}
\providecommand\citep{\citep}

\newif\iflatexml\latexmlfalse
\AtBeginDocument{\DeclareGraphicsExtensions{.pdf,.PDF,.eps,.EPS,.png,.PNG,.tif,.TIF,.jpg,.JPG,.jpeg,.JPEG}}

\usepackage[utf8]{inputenc}
\usepackage[english]{babel}

\begin{document}

\title{Sex differences in network controllability as a predictor of executive function in youth}

\author[1,2]{Eli J. Cornblath}
\author[2]{Evelyn Tang}
\author[1,2,3]{Graham L. Baum}
\author[3]{Tyler M. Moore}
\author[3]{David R. Roalf}
\author[3]{Ruben C. Gur}
\author[3]{Raquel E. Gur}
\author[4]{Fabio Pasqualetti}
\author[3,*]{Theodore D. Satterthwaite}
\author[2,5,6,7,*]{Danielle S. Bassett}
\affil[1]{Department of Neuroscience, University of Pennsylvania, Philadelphia, PA 19104 USA}
\affil[2]{Department of Bioengineering, University of Pennsylvania, Philadelphia, PA 19104 USA}
\affil[3]{Department of Psychiatry, University of Pennsylvania, Philadelphia, PA 19104 USA}
\affil[4]{Department of Mechanical Engineering, University of California, Riverside, CA 92521 USA}
\affil[5]{Department of Electrical and Systems Engineering, University of Pennsylvania, Philadelphia, PA 19104 USA}
\affil[6]{Department of Neurology, University of Pennsylvania, Philadelphia, PA 19104 USA}
\affil[7]{To whom correspondence should be addressed: dsb@seas.upenn.edu.}
\affil[*]{These authors contributed equally.}

\vspace{-1em}

  \date{\today}

\begingroup
\let\center\flushleft
\let\endcenter\endflushleft
\maketitle
\endgroup

\newpage
\selectlanguage{english}
\begin{abstract}
\normalsize{Executive function is a quintessential human capacity that emerges late in development and displays different developmental trends in males and females. Sex differences in executive function in youth have been linked to vulnerability to psychopathology as well as to behaviors that impinge on health, wellbeing, and longevity. Yet, the neurobiological basis of these differences is not well understood, in part due to the spatiotemporal complexity inherent in patterns of brain network maturation supporting executive function. Here we test the hypothesis that sex differences in executive function in youth stem from sex differences in the controllability of structural brain networks as they rewire over development. Combining methods from network neuroscience and network control theory, we characterize the network control properties of structural brain networks estimated from diffusion imaging data acquired in males and females in a sample of 882 youth aged 8-22 years. We summarize the control properties of these networks by estimating average and modal controllability, two statistics that probe the ease with which brain areas can drive the network towards easy- versus difficult-to-reach states. We find that females have higher modal controllability in frontal, parietal, and subcortical regions while males have higher average controllability in frontal and subcortical regions. Furthermore, average controllability values in the medial frontal cortex and subcortex, both higher in males, are negatively related to executive function. Finally, we find that average controllability predicts sex-dependent individual differences in activation during an n-back working memory task. Taken together, our findings support the notion that sex differences in the controllability of structural brain networks can partially explain sex differences in executive function. Controllability of structural brain networks also predicts features of task-relevant activation, suggesting the potential for controllability to represent context-specific constraints on network state more generally.}
\end{abstract}%

\flushbottom
\thispagestyle{empty}

\section*{Keywords:} network controllability, neurodevelopment, sex differences, executive function, working memory, fMRI BOLD, diffusion tensor imaging

\newpage

\section*{Introduction}

Executive function is necessary for regulation of goal-directed behavior, and encompasses cognitive processes including working memory, inhibition, task switching, and performance monitoring \citep{anderson2001development}. Deficits in executive function are associated with increased risk taking and associated consequences \citep{ROMER20092916, barkley2002driving}, and more generally hamper academic and occupational performance \citep{bied}. Executive deficits frequently lead to personal, social, and professional consequences that accumulate throughout the course of a patient's life \citep{bied}. Importantly, the normative capacity for executive function rapidly increases during adolescence and differs by sex. Sex differences in the developmental trajectory of executive function have been linked to higher rates of impulsivity \citep{chap}, ADHD diagnosis \citep{willcutt2012prevalence}, criminality \citep{cross2011sex} and substance use \citep{ROMER20092916} in males. Current interventions for disorders of executive function are relatively limited, usually do not consider sex, and rely primarily on psychotherapy and global manipulations via psychopharmacology \citep{hosenbocus2012review}. 

A basic understanding of sex-related differences in executive function and their implications for the diagnosis and treatment of executive deficits requires an understanding of the normative maturation of underlying neural circuitry. Several recent studies highlight the fact that such maturation, and sex-differences in that maturation, span structure \cite{gogtay2004dynamic}, anatomical connectivity \cite{baum2017modular}, functional activity \cite{schmithorst2015evidence,nomi2017moment,keulers2011developmental}, and functional connectivity \cite{fair2009functional}. Using structural MRI, a recent study \citep{Gennatas3550-16} reported age-related, non-linear increases in gray matter density with concurrent decreases in cortical thickness. Interestingly, the maturation of these structural features was markedly different between the sexes, with females showing higher gray matter density globally and higher cortical thickness in frontal and insular regions \citep{Gennatas3550-16}. Several older structural studies \citep{blake} have also reported linear increases in frontoparietal white matter density throughout adolescence. Using diffusion-weighted MRI, another study found significantly greater within-hemisphere connectivity in males and greater between-hemisphere connectivity in females \citep{sattconnsex2014}. Sex differences have also been identified in the clustered (or modular) structure in patterns of functional connectivity estimated from resting state fMRI data: males display higher between-module connectivity while females display higher within-module connectivity \citep{satterthwaite2014linked}, and these differences were shown to predict individual differences in executive function \citep{satterthwaite2014linked}. Although these descriptive studies have provided important insights, it remains difficult to specify in a mechanistic sense how executive function might arise from such complex, multimodal patterns of brain maturation in a sex-dependent manner.

\begin{figure}[h]
\begin{center}
\includegraphics[width=18cm,keepaspectratio]{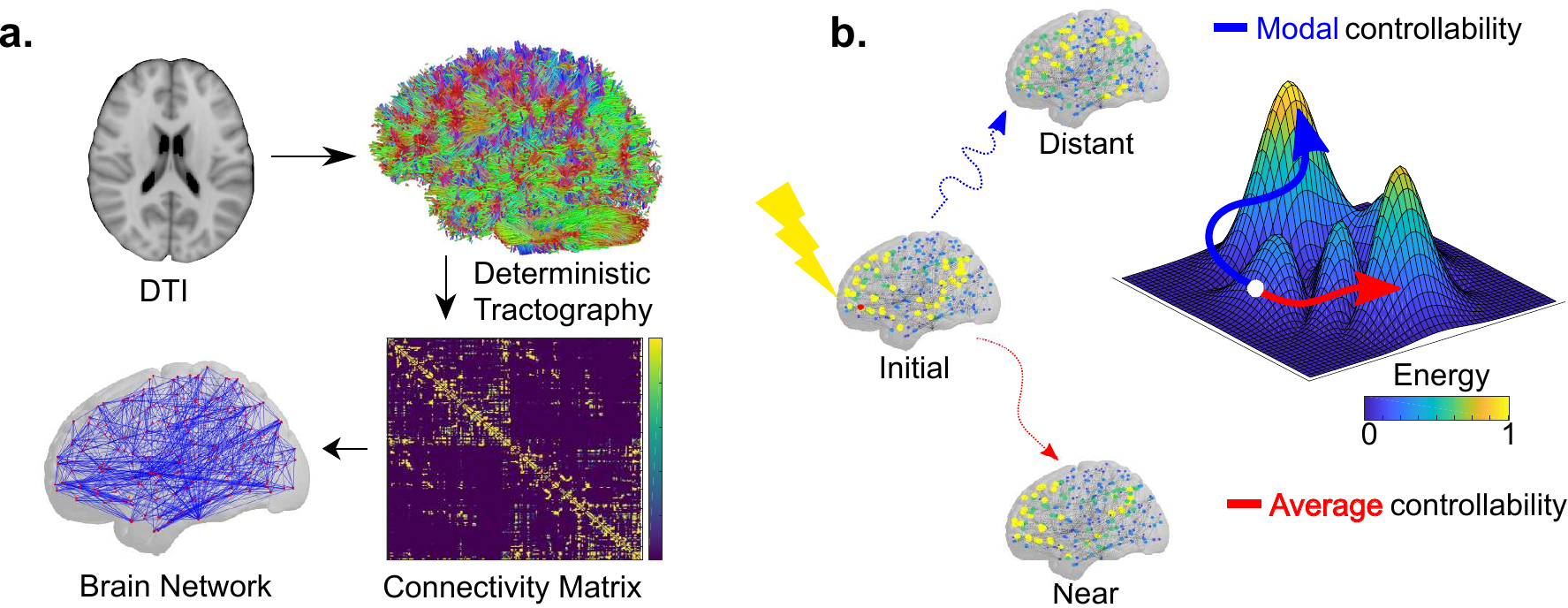}
\caption{{\textbf{Control theory and schematic of data processing.} \emph{(a)} Schematic depicting basic methodological approach. Diffusion tensor imaging (DTI) data was acquired from 882 youth ages 8 to 22 years. Deterministic tractography was used to identify the number of white matter streamlines, connecting any two regions of interest. These estimates were used to construct a structural brain network for each subject representing white matter connectivity (edges) between brain regions (nodes).  \emph{(b)} Diagram illustrating brain state transitions from an activated default mode system (\textit{Initial}). The blue arrow denotes a distant transition to a deactivated default mode system and activated frontoparietal/dorsal attention system (\textit{Distant}). The red arrow denotes a nearby transition to a partially deactivated default mode system (\textit{Near}). Regions with high modal controllability can facilitate transitions to energetically distant states while regions with high average controllability can facilitate transitions to nearby states while requiring very little energy.
\label{fig:figure1}%
}}
\end{center}
\end{figure}

We address this challenge by positing that sex differences in executive function in youth stem from sex differences in the controllability of structural brain networks as they rewire over development. This notion intuitively bridges the control of behavior (executive function) with the control of brain dynamics (network controllability). Specifically, we capitalize on recent advances in network control theory \cite{liu2011controllability,pasqualetti2014controllability}, an emerging branch of theoretical physics and systems engineering that builds on early efforts in control theory \citep{TK:80,kalman1963} to offer a mechanistic model of how key nodes, or control points, can exert disproportionate influence over system function \cite{liu2011controllability}. Control points are identified with metrics that assess the ability of specific nodes to alter a system’s state, based on the underlying network topology \cite{pasqualetti2014controllability} (Fig. \ref{fig:figure1}). Specifically, the metric of \emph{average controllability} reflects the average energy input required at a node to move the system from some initial state to all possible states. In contrast, the metric of \emph{modal controllability} reflects the ease of transitioning the system from some initial state to a difficult-to-reach state. Prior work has demonstrated the utility of network control theory in understanding basic brain architecture and function \cite{muldoon2016stimulation,betzel2016optimally,gu2017optimal} across spatial scales \cite{wiles2017autaptic} and species \cite{kim2017role}, posited its relation to cognition \cite{gu2015controllability,tang2017developmental}, and outlined its developmental course \cite{tang2017developmental}. 

Here, we tested the hypothesis that developmental sex differences in network control underlie sex differences in executive functioning. Specifically, we predicted that (i) network controllability differs by sex, (ii) network controllability changes with age differently in males and females, (iii) sex differences in network controllability predict executive function, and (iv) network controllability predicts the activation of brain regions during a working memory task demanding executive function. To test these hypotheses, we constructed structural brain networks from diffusion tensor imaging data acquired in 882 healthy youth, ages 8-22 years, in the Philadelphia Neurodevelopmental Cohort (PNC) \citep{satterthwaite2014neuroimaging}. Each brain network was comprised of 234 anatomically defined brain regions \cite{cammoun2012mapping} connected by white matter tracts estimated from diffusion tractography. We show that regional controllability is a significant mediator of the relationship between sex and executive function, and that it predicts the magnitude of fMRI BOLD signal on an n-back working memory task. As described in detail below, our results suggest that sex differences in the controllability of structural brain networks predict executive function and the activity profiles that support that function.

\section*{Materials and Methods}

\subsection*{Participants}

Diffusion tensor imaging (DTI) data were obtained from youth who participated in a large community-based study of brain development, now known as the Philadelphia Neurodevelopmental Cohort (PNC) \citep{satterthwaite2014neuroimaging}. Here we study 882 out of a total of 1601 subjects between the ages of 8 and 22 years (mean age = 15.06, SD = 3.15, 389 males, 493 females). Due to lack of complete diffusion scans ($n = 224$) and incidental findings ($n=20$), data from 244 participants was deemed unusable. The remaining 1357 participants underwent a rigorous manual and automated quality assurance protocol for DTI datasets \citep{roalf2016impact}, eliminating an additional 147 subjects with poor data quality. A subset of 93 of the remaining 1210 participants were excluded for low quality or incomplete FreeSurfer reconstruction of T1-weighted images. Further, 235 of the remaining 1117 participants were excluded for one or more of the following reasons: gross radiological abnormalities distorting brain anatomy, medical history that might impact brain function, history of inpatient psychiatric hospitalization, use of psychotropic medication at the time of imaging, or high levels of in-scanner head motion during the DTI scan, as defined by a mean relative displacement between non-weighted volumes of greater than 2 mm. These exclusions left us with a final sample of $n=882$ subjects \citep{baum2017modular,tang2017developmental} between the ages of 8 and 22 years (mean age = 15.06, SD = 3.15, 389 males, 493 females). 

\subsection*{Cognitive Phenotyping}

Cognition was measured outside of the scanner using the Penn Computerized Neurocognitive Battery (CNB) \citep{gur2010cognitive,gur2012age}. Briefly, the 1-hour CNB was administered to all participants, and consisted of 14 tests that evaluated a broad range of cognitive functions. Twelve of the tests measure both accuracy and speed, while two of the tests (motor and sensorimotor) measure only speed.

Overall cognitive performance was summarized as the average $z$-transformed accuracy and speed (as well as the difference between accuracy and median response time: efficiency) scores across all tests administered (as described in Moore et al.\citep{moore2015psychometric} for complete details). Factor scores are described in Moore et al.\citep{moore2015psychometric}; here, we used the factor score for executive efficiency from a best-fitting four-factor solution comprising tests from the executive function domain (attention, abstraction and working memory). The tests contributing to the executive efficiency score include the Penn Continuous Performance Test, the Letter N-Back task, and (weakly) the Penn Verbal Reasoning Test \citep{moore2015psychometric,moore2016development, moore2017development} (analogical reasoning). In the present study, we use this factor score for executive efficiency as our primary measure of executive function, hereafter referred to as ``executive function''.

\subsection*{Imaging Data Acquisition}

MRI data were acquired on a 3 Tesla Siemens Tim Trio whole-body scanner and 32-channel head coil at the Hospital of the University of Pennsylvania. DTI scans were acquired via a twice-refocused spin-echo (TRSE) single-shot echo-planar imaging (EPI) sequence (TR = 8100ms, TE = 82ms, FOV = 240mm$^2$/240mm$^2$; Matrix = RL:128/AP:128/Slices:70, in-plane resolution (x and y) 1.875 mm$^2$; slice thickness = 2mm, gap = 0; flip angle = 90/180/180 degrees, volumes = 71, GRAPPA factor = 3, bandwidth = 2170 Hz/pixel, PE direction = AP). This sequence utilizes a four-lobed diffusion encoding gradient scheme combined with a 90-180-180 spin-echo sequence designed to minimize eddy-current artifacts. The complete sequence consisted of 64 diffusion-weighted directions with $b$ = 1000s/mm$^2$ and 7 interspersed scans where $b$ = 0s/mm$^2$. Total scan time was approximately 11 min. The imaging volume was prescribed in axial orientation covering the entire cerebrum with the topmost slice just superior to the apex of the brain. 

In addition to the DTI scan, a map of the main magnetic field (i.e., B0) was derived from a double-echo, gradient-recalled echo (GRE) sequence, allowing us to estimate field distortions in each dataset. Prior to DTI acquisition, a 5-minute magnetization-prepared, rapid acquisition gradient-echo T1-weighted (MPRAGE) image (TR 1810 ms, TE 3.51 ms, FOV 180 $\times$240 mm, matrix 256 $\times$ 192, effective voxel resolution of 1 $\times$ 1 $\times$ 1 mm) was acquired. This high-resolution structural image was used for tissue segmentation and parcellating gray matter into anatomically defined regions in native space. Rigorous manual and automated quality-assurance protocols for the T1-weighted structural imaging data were performed and cleared for the 882 subjects considered here \citep{vandekar2015topologically}. Subsequently, all structural images were processed using FreeSurfer (version 5.3) \cite{fischl2012freesurfer}. FreeSurfer reconstructions underwent rigorous quality-assurance protocols \citep{vandekar2015topologically,Rosen2017}. The T1 image was parcellated into 234 regions by FreeSurfer according to the Lausanne Atlas \citep{lausanne}. We define the subcortex of this 234-region parcellation to be comprised of the left and right hemispheric counterparts of the thalamus proper, caudate, putamen, pallidum, nucleus accumbens area, hippocampus and amygdala, while excluding the brainstem (14 regions). We define the cortex of this 234-region parcellation to be comprised of the remaining regions (219 regions).

fMRI BOLD data were acquired as subjects completed a version of the n-back task using fractal images \citep{ragland2002working}. See the supplementary methods of ref. \citep{shanmugan2016common} for details regarding task presentation and structure. Functional images were obtained using a whole-brain, single-shot, multislice, gradient-echo echoplanar sequence (231 volumes; TR $= 3000$ ms; TE $= 32$ ms; flip angle $= 90$ degrees; FOV = $192 \times 192$ mm; matrix = $64 \times 64$; slices $= 46$; slice thickness $= 3$ mm; slice gap $= 0$ mm; effective voxel resolution = $3.0 \times 3.0 \times 3.0$ mm). 

\subsection*{Imaging Data Preprocessing}

All DTI datasets were subject to a rigorous manual quality assessment procedure involving visual inspection of all 71 volumes \citep{roalf2016impact}. Each volume was evaluated for the presence of artifact, with the total number of volumes impacted summed over the series. This scoring was based on previous work describing the impact of removing image volumes when estimating the diffusion tensor
\cite{jones2004squashing,chen2015effects}. Data was considered ``Poor'' if more than 14 (20\%) volumes contained artifact, ``Good'' if it contained 1-14 volumes with artifact, and ``Excellent'' if no visible artifacts were detected in any volumes. All 882 subjects included in the present study had diffusion datasets identified as ``Good'' or ``Excellent,'' and had less than 2mm mean relative displacement between interspersed $b = 0$ volumes. As described below, even after this rigorous quality assurance, motion was included as a covariate in all analyses.

The skull was removed for each subject by registering a binary mask of a standard fractional anisotropy (FA) map (FMRIB58 FA) to each subject’s DTI image using an affine transformation \cite{jenkinson2002improved}. Eddy currents and subject motion were estimated and corrected using the FSL eddy tool \cite{andersson2016integrated}. Diffusion gradient vectors were then rotated to adjust for subject motion estimated by eddy. After the field map was estimated, distortion correction was applied to DTI data using FSL’s FUGUE \cite{jenkinson2012fsl}.

BOLD time series were processed as described in \citep{satterthwaite2013functional,shanmugan2016common}. Briefly, FSL 5 \citep{jenkinson2012fsl} was used to analyze time series data from three condition blocks (0-back, 1-back and 2-back), with the primary contrast being 2-back $>$ 0-back. BOLD images were co-registered to the T1 image using boundary-based registration \citep{greve2009accurate} with integrated distortion correction as implemented in FSL. Generalized linear model (GLM) beta weights were averaged across all voxels in each parcel of the 234-node Lausanne atlas. In our assessment of n-back performance-related activation, we use the difference in GLM beta weights between the 2-back and 0-back condition. For all analyses of fMRI data, we excluded 223 subjects with incomplete data or excessive head motion (mean relative displacement $> 0.5$ mm or maximum displacement $> 6$ mm), leaving $n = 659$ remaining.

\subsection*{Structural Network Estimation} 

Structural connectivity was estimated from DTI data in order to generate the adjacency matrix representing the pattern of white matter tracts between large-scale brain areas. DSI Studio was used to estimate the diffusion tensor and perform deterministic whole-brain fiber tracking with a modified FACT algorithm that used exactly 1,000,000 streamlines per subject after removing all streamline with length $< 10$ mm \citep{gu2015controllability}. To extend regions into white matter, parcels defined using the Lausanne atlas were dilated by 4 mm \citep{gu2015controllability,tang2017developmental} and registered to the first non-weighted (\textit{b} = 0) volume using an affine transform \citep{gu2015controllability,tang2017developmental}. The number of streamlines connecting each node of the 234 region parcel end-to-end was used to define the edge weights $a_{ij}$ of the adjacency matrix $A$. 
\label{NetworkConstruction}

\subsection*{Network Controllability}
\label{NetworkControllability}
We represent the streamline-weighted structural network estimated from diffusion tractography as the graph $\mathcal{G} = (\mathcal{V}, \mathcal{E})$, where $\mathcal{V}$ and $\mathcal{E}$ are the vertex and edge sets, respectively. Let $a_{ij}$ be the weight associated with the edge $(i, j) \in \mathcal{E}$, and define the weighted adjacency matrix of $\mathcal{G}$ as $A = [a_{ij}]$, where $a_{ij} = 0$ whenever $(i, j)\notin\mathcal{E}$. We associate a real value with each node to generate a vector describing the network state, and we define the map $x : \mathbb{N}_{\geq0} \leftarrow \mathbb{R}^{N}$ to describe the dynamics of the network state over time. 

It is worth noting that this method assumes that the number of streamlines is proportional to the strength of structural connectivity with regards to propagation of activity between nodes according to a specified model of dynamics. Here we employ a simplified noise-free linear discrete-time and time-invariant model of such dynamics:

\begin{align}
\mathbf{x}(t + 1) = \mathbf{Ax}(t) + \mathbf{B}_{\mathcal{K}}\mathbf{u}_{\mathcal{K}}(t),
\end{align}\label{eq:cont}

where $\mathbf{x}$ describes the state (i.e. voltage, firing rate, BOLD signal) of brain regions over time. Thus, the state vector $\mathbf{x}$ has length $N$, where $N$ is the number of brain regions in the connectome parcellation, and the value of $\mathbf{x}_i$ describes the brain activity state of that region. The matrix $\mathbf{A}$ is symmetric, with the diagonal elements satisfying $A_{ii} = 0$. Prior to calculating controllability values, we divide $\mathbf{A}$ by $1 + \xi_0(\mathbf{A})$, where $\xi_0(\mathbf{A})$ is the largest eigenvalue of $\mathbf{A}$. The input matrix $\mathbf{B}_\mathcal{K}$ identifies the control point $\mathcal{K}$ in the brain, where $\mathcal{K} = {k_1,...,k_m}$ and

\begin{align}
\mathbf{B}_{\mathcal{K}} = [e_{k_{1}}  \cdots e_{k_{m}}],
\end{align}
 and $e_i$ denotes the $i$-th canonical vector of dimension $N$. The input $\mathbf{u}_{\mathcal{K}} : \mathbb{R}_{\geq0}\rightarrow\mathbb{R}^M$ denotes the control strategy.

To study the dynamics by which the activity of one brain region influences structurally connected regions, we apply the control theoretic notion of controllability to our dynamical model. Classic results in control theory ensure that controllability of the network, $\mathbf{x}(t + 1) = \mathbf{Ax}(t) + \mathbf{B}_{\mathcal{K}}\mathbf{u}_{\mathcal{K}}(t)$, from the set of network nodes $\mathcal{K}$ is equivalent to the controllability Gramian $\mathbf{W}_{\mathcal{K}}$ being invertible, where

\begin{align}
\mathbf{W}_{\mathcal{K}} = \sum_{\tau = 0}^\infty\mathbf{A}^\tau\mathbf{B}_{\mathcal{K}}\mathbf{B}_{\mathcal{K}}^T\mathbf{A}^\tau.
\end{align}

We calculate $\mathbf{W}_{\mathcal{K}}$, or rather $\mathbf{W}_i$, with $\mathbf{B}_{\mathcal{K}}$ set equal to one canonical vector $e_i$ and repeat this process for all $N$ nodes \citep{tang2017developmental,gu2015controllability}. Although it is well known that the activity of several brain regions and neuronal ensembles are related via non-linear dynamics, it has been shown that a linear approximation can explain features of the resting state fMRI BOLD signal \citep{honey2009predicting}; this suggests that a linear approximation can effectively capture the controllability properties of the original non-linear dynamics.

\subsection*{Controllability Metrics}

Following ref.\citep{tang2017developmental,gu2015controllability}, controllability metrics for structural brain networks were calculated for two different control strategies which describe the ability to change $\mathbf{x}(t)$ in a particular fashion \citep{pasqualetti2014controllability}. Average controllability describes the ease of transition to energetically similar states, while modal controllability describes the ease of transition to difficult-to-reach states \citep{pasqualetti2014controllability}.

Average controllability of a network equals the average input energy applied to a set of control nodes required to reach all possible target states. It is known that average input energy is proportional to Trace($\mathbf{W}_{\mathcal{K}}^{-1}$), the trace of the inverse of the controllability Gramian. Instead, as in ref. \citep{tang2017developmental,gu2015controllability}, we use Trace($\mathbf{W}_{\mathcal{K}}$) as a measure of average controllability because (i), Trace($\mathbf{W}_{\mathcal{K}}$) and Trace($\mathbf{W}_{\mathcal{K}}^{-1}$) are related via inverse proportionality, and (ii), Trace($\mathbf{W}_{\mathcal{K}}^{-1}$) tends to be very ill-conditioned and cannot be accurately computed even at more coarse connectome parcellations. In addition to its relationship with Trace($\mathbf{W}_{\mathcal{K}}^{-1}$), Trace($\mathbf{W}_{\mathcal{K}}$) describes the energy of the network impulse response, or, equivalently, the network $H_2$ norm \cite{TK:80,kalman1963}. In summary, to compute the average controllability value for node $i$ in $\mathbf{A}$, we compute the Trace($\mathbf{W}_{\mathcal{K}}$) when node $i$ is the only control node (i.e. $\mathbf{B}_{\mathcal{K}} = e_i$).

Modal controllability refers to the ability of a node to control the evolutionary modes of a dynamical network, and is most interpretable when used to identify states that are poorly controllable given $\mathbf{B}_{\mathcal{K}}$. To calculate modal controllability, one must first obtain the eigenvector matrix $V = [v_{ij}]$ of the adjacency matrix $\mathbf{A}$. If $v_{ij}$ is small, then the $j$-th evolutionary mode of the input-independent form of Eq. (1), $\mathbf{x}(t) = \mathbf{Ax}(t)$, is poorly controllable from node $i$. According to prior work \citep{pasqualetti2014controllability}, we define $\phi_i = \sum_{j=1}^{N}(1 - \xi_{j}^{2}(\mathbf{A}))v_{ij}^{2}$ as a scaled measure of the modal controllability of each of the $N$ modes $\xi_0(\mathbf{A}), ..., \xi_{N-1}(\mathbf{A})$ from node $i$.

Finally, we calculate the mean of the regional controllability values either across the whole brain (Fig. \ref{fig:figureS1}) or within the cortex and subcortex separately (Figs. \ref{fig:figure2}, \ref{fig:figure3}, \ref{fig:figure4}, \ref{fig:figure5}). For each subject, we define the mean modal controllability to be the sum of the values of $\phi_i$ for each node within $\mathbf{A}$, divided by the number of regions. Similarly, we define the mean average controllability to be the sum of the values of Trace($\mathbf{W}_{\mathcal{K}}$) for each node within $\mathbf{A}$, divided by the number of regions. 

\subsection*{Network Synchronizability} 

While the metrics of average and modal controllability provide complementary views on the diverse dynamics that a brain network can display, it is also useful to study contrasting metrics that probe the susceptibility of the network to be constrained within a narrower range of dynamics. Previous work \citep{tang2017developmental} has demonstrated that a useful contrasting metric to controllability is synchronizability, which intuitively measures the susceptibility of a network to remain in a single synchronous state $\mathbf{s}(t)$, i.e. $\mathbf{x}_1 = \cdots = \mathbf{x}_n(t+1) = \mathbf{s}(t)$. The master stability function (MSF) allows one to analyze the stability of this synchronous state without fully characterizing the properties of each feature of the system \citep{pecora1998master}. Within this framework, linear stability depends on the positive eigenvalues ${\lambda_i}, i=1, ..., N-1$ of the Laplacian matrix $\mathbf{L}$ defined by $L_{ij}=\delta_{ij}\sum_k A_{ik}-A_{ij}$, where $\delta_{ij}$ is the Kronecker delta function.

The condition for stability depends on the shape of the MSF, i.e. the
stability of the synchronized state to linear perturbations holds when
the MSF is negative for all non-zero eigenvalues of $\mathbf{L}$. This
condition is more likely for a smaller range of eigenvalues, hence we
can use the normalized spread of these eigenvalues to quantify network
synchronizability as

\begin{align}
\frac{1}{\sigma^{2}} = \frac{d^{2}(N-1)}{\sum_{i = 1}^{N-1}|\lambda_{i} - \bar{\lambda}|^2}, ~~\mathrm{where}~~ \bar{\lambda} := \frac{1}{N-1}\sum_{i=1}^{N-1}\lambda_i
\end{align}

and $d := \frac{1}{N} \sum_{i} \sum_{j\neq i} A_{ij}$, the average coupling strength per node, which normalizes for the overall network strength. Using this definition, we calculated the synchronizability of $\mathbf{A}$ for each subject. 

\subsection*{Linear Regression of Network Control Metrics on Sex} 

For all analyses of network control metrics, we examined the effect of sex while controlling for age, total brain volume (segmented brain volume, as defined by FreeSurfer BrainSegVol metric), handedness, and motion during the diffusion scan. We used multiple ordinary least squares (OLS) linear regression with the $lm()$ command in R or the MATLAB $regress$ function to fit the following general equation:

\begin{equation} \label{eq:equation5}
C = 1 + \beta_{a} a + \beta_{v} v + \beta_{h} h + \beta_{m_{d}} m_{d} + \beta_{s} s~, 
\end{equation} 

where $C$ is the controllability statistic (either $\phi_{i}$ for modal controllability or $\mathcal{A}$ for average controllability), $a$ is age, $v$ is total intracranial volume, $m_{d}$ is the mean framewise displacement as a summary measure of in-scanner head motion during the diffusion imaging sequence, $h$ is handedness, and $s$ is sex. We then used the R package \textbf{visreg} to calculate 95\% confidence intervals around fitted lines and generate partial residuals. Furthermore, when using the multiple OLS regression for the node-level analyses of controllability, we applied a false discovery rate (FDR) correction \citep{bhfdr} ($q < 0.05$) to control for Type I error due to multiple testing.

\subsection*{Non-Linear Fits of Executive Function and Network Metrics}

Because age and executive function are nonlinearly related, we used a general additive model \citep{wood2004stable,vandekar2015topologically} (GAM) to assess the relationship between executive efficiency and age using the $mgcv$ package in R (Fig. \ref{fig:figure1}b). A GAM is a generalized linear model in which the linear predictor is defined by unknown smooth functions of predictor variables. In this case, we use penalized regression splines for age, with conventional parametric components for the remaining predictors.

Following ref. \citep{tang2017developmental}, we used non-linear models to examine the relationships between synchronizability, average controllability, and modal controllability (Fig. \ref{fig:figure4}). We generated estimates of parameters for models of the form $y = a + b\exp(cx)$ via non-linear least squares using the $nls()$ function in R. To compare these fits between males and females, we performed an analysis of variance (ANOVA). For example, when considering sex differences in the nonlinear relation between modal controllability and average controllability, we examined the expression:

\begin{equation}\label{eq:equation6}
\phi_{i} = (a + \alpha s) + (b + \beta s)\exp((c+\gamma s)\mathcal{A})~,
\end{equation}

where $\phi_{i}$ is modal controllability, $\mathcal{A}$ is average controllability, and $s$ is sex, coded as 0 for males and 1 for females, so that $\alpha$, $\beta$ and $\gamma =0$ for males and the equation is reduced to the base form. The predicted values for males or females obtained from the full model were used to generate the curves in Fig. \ref{fig:figure4}.
 
\subsection*{Mediation Testing}

After identifying associations between sex, controllability, and executive function, we asked whether controllability is a statistically significant mediator of the relationship between sex and executive function. To test for a formal mediation, we performed non-parametric bootstrapping causal mediation analysis using the R function $mediate()$ from the package \textbf{mediation} \citep{tingley2014mediation}. It is worth noting that we cannot estimate causality in cross-sectional association data. The Average Causal Mediation Effect (ACME) quantifies the relationship between mediator and outcome independent of treatment, while the Average Direct Effect (ADE) quantifies the relationship between treatment and outcome, independent of the mediator \citep{tingley2014mediation}.

\subsection*{Linear Regression of BOLD Data on Network Control Metrics}

Finally, and in keeping with our linear systems framework, we sought to test whether controllability would predict brain state ($\mathbf{x}(t)$), as defined by BOLD activation during a task demanding executive function. In these analyses (Fig. \ref{fig:figure7}), we used the residuals from Eq.~\ref{eq:equation5} for controllability. For activation, we used the residuals from the following equation: 

\begin{equation} \label{eq:equation5.5}
A = 1 + \beta_{a} a + \beta_{h} h + \beta_{m_{n}} m_{n} + \beta_{s} s~,
\end{equation}

where $A$ is 2-back minus 0-back activation (hereafter referred to as ``activation''), $a$ is age, $h$ is handedness, $m_{n}$ is the mean framewise displacement during the n-back scan, and $s$ is sex. Because we had no prior knowledge about where controllability might predict activation, we fit linear models between controllability and activation at every possible pair of nodes, i.e. we performed $234 \times 234$ regressions of controllability residuals at each node with activation residuals at each node. The $234 \times 234$ matrices of $p$-values for the slope of controllability on activation were FDR corrected ($q<0.05$) separately for average controllability and for modal controllability. To identify regions where activation was associated with executive function, we followed previous work \citep{satterthwaite2013functional,shanmugan2016common} by examining $d^\prime$, a composite measure of n-back task performance which takes both correct responses and false positives into account to separate performance from response bias. 

\section*{Results}

\subsection*{Sex differences in executive function}

\begin{figure}[h]
	\begin{center}
		\includegraphics[width=18cm,keepaspectratio]{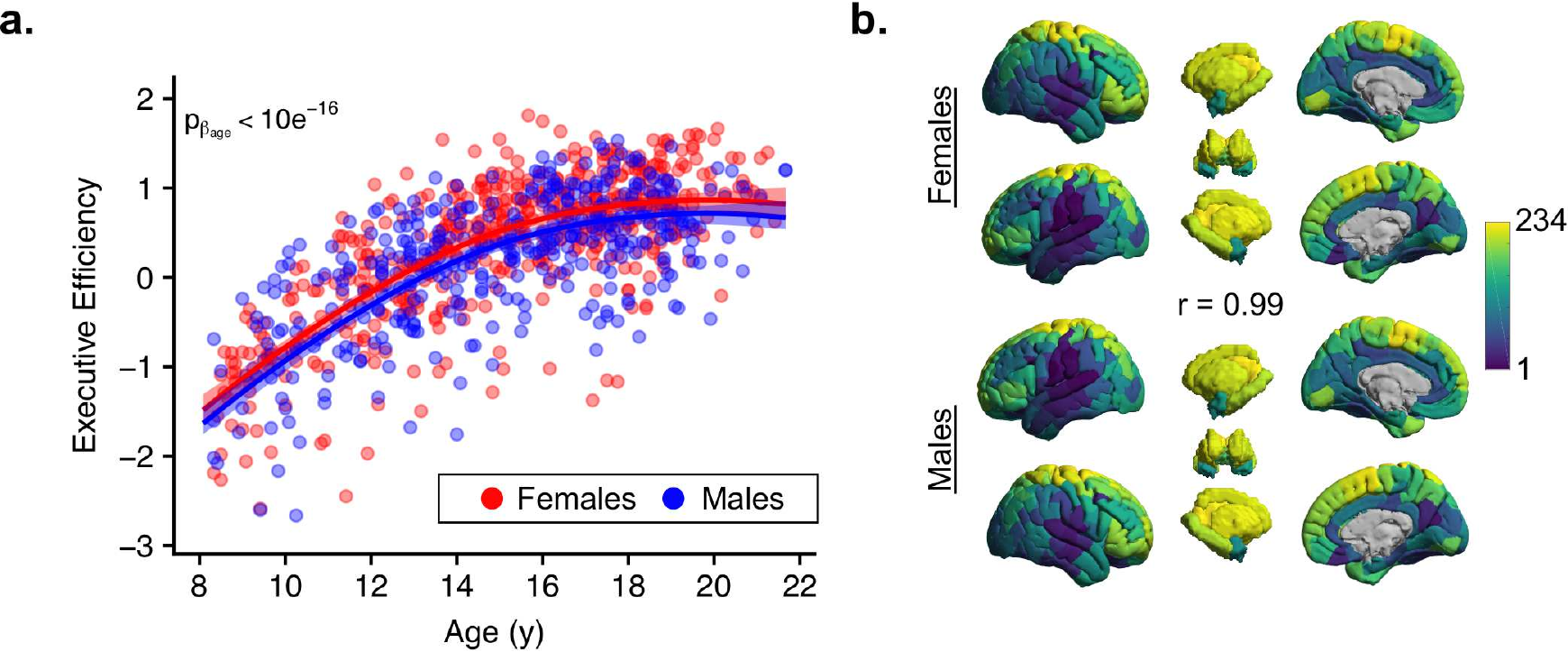}
		\caption{{\textbf{Cognitive development and network controllability by sex.} \emph{(a)} Executive efficiency, a summary of performance on tasks taxing executive function in the PCNB, fit non-linearly on age with a penalized regression spline. Red and blue data points and curves represent females and males, respectively. Each point represents a subject's partial residual with respect to sex; solid lines represent the predicted mean for males and females separately, and shaded envelopes denote 95\% confidence intervals of the prediction. \emph{(b)} Rank of mean average controllability values at each node across 389 male subjects and 493 female subjects, separately. The Spearman rank correlation between regional average controllability averaged across males and regional average controllability averaged across females was $r = 0.9915$.
				\label{fig:figure1_5}%
		}}
	\end{center}
\end{figure}

Our analysis of sex differences in the development of executive function and its neurobiological underpinnings began with a sex-stratified comparison of mean performance on \textit{executive efficiency}, a cross-contextual summary measurement of executive function \citep{moore2015psychometric}. Consistent with prior results from the PNC data \citep{satterthwaite2014linked, satterthwaite2014neuroimaging}, the executive efficiency subscore of the PCNB was significantly higher in females (full model $r^2 = 0.52$, $df = 873$, $p = 0.0015$; Fig. \ref{fig:figure1_5}a) with no interaction between age and sex. Furthermore, executive efficiency increased with age in a non-linear fashion ($r^2 = 0.52$, $df = 2$, $p < 10^{-15}$; Fig. \ref{fig:figure1_5}a). This result suggests that development of executive function occurs via a similar course for males and females, but that females display higher scores in this cognitive domain for all ages studied. After performing this sex-stratified analysis of the developmental course of the executive efficiency score, we next turned to a consideration of its neurobiological underpinnings (Fig. \ref{fig:figure1}a).

\subsection*{Sex differences in regional network controllability}

Our general hypothesis was that sex differences in executive function in youth stem from sex differences in the controllability of structural brain networks as they rewire over development, a notion that bridges the control of behavior (executive function) with the control of brain dynamics (network controllability). To test this hypothesis, we examined the anatomical distribution of control points in the structural brain networks of males and females separately. We ranked the mean average controllability value at each region for males and females separately, which revealed that the distribution of controller strength is virtually identical between male and females ($r = 0.99$; Fig. \ref{fig:figure1_5}b). This result suggests that there is no sex difference in the spatial distribution of controllers when classified by their relative magnitudes. 

\begin{figure}[h!]
\begin{center}
\includegraphics[width=18cm,keepaspectratio]{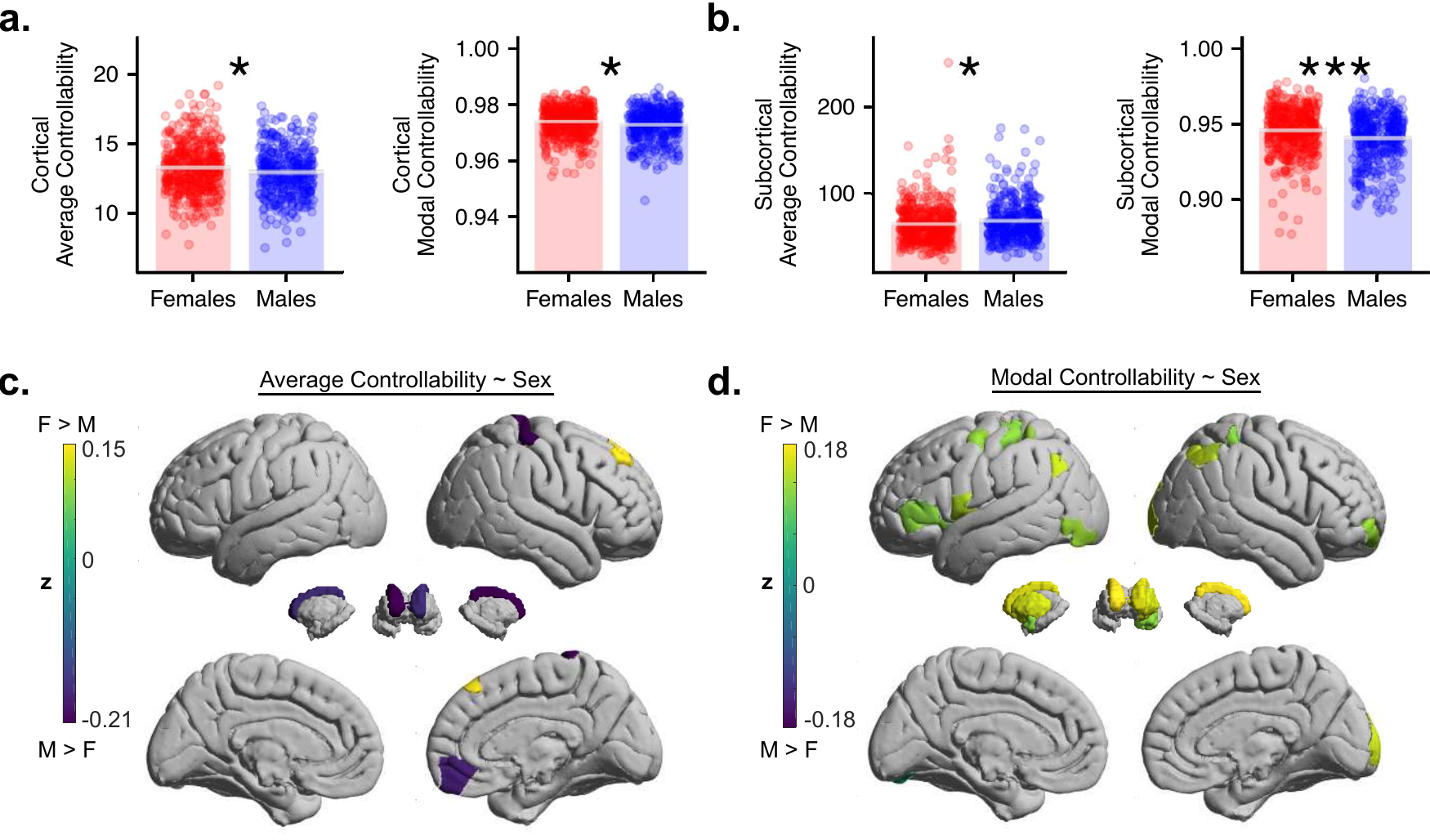}
\caption{{\textbf{Sex differences in regional controllability.} \emph{(a-b)} Controllability estimates averaged over 219 cortical nodes \emph{(a)} and 14 subcortical nodes \emph{(b)}, for average controllability and modal controllability. Females have higher cortical average controllability, cortical modal controllabillity and subcortical modal controllability, but males have higher subcortical average controllability. In panels \emph{a-b}, red and blue represent females and males, respectively. Each point represents a subject's partial residual with respect to sex; bar height represents the predicted mean for males and females separately, and shaded grey envelopes denote 95\% confidence intervals of the prediction. \emph{(c-d)} Heatmaps where color intensity corresponds to standardized regression coefficient for sex, with warmer colors indicating higher controllability in females. Average controllability is depicted in \emph{(c)} and modal controllability is depicted in \emph{(d)}. Colors reflect values of standardized regression coefficients for controllability at each node as a predictor of executive function while controlling for covariates (Eq. \ref{eq:equation5}). *, $p < 0.05$, **, $p < 0.01$, ***, $p < 10^{-4}$.
\label{fig:figure2}%
}}
\end{center}
\end{figure}

We next investigated whether the actual (rather than ranked) regional controllability estimates differed by sex, and we addressed this question first by considering the cortex and subcortex separately. Specifically, we computed mean controllability values for each subject across 219 cortical regions and 14 subcortical regions. In the cortex, mean average controllability ($r^2 = 0.082$, $df = 876$, $p = 0.018$; Fig. \ref{fig:figure2}a) and mean modal controllability ($r^2 = 0.13$, $df = 876$, $p = 0.018$; Fig. \ref{fig:figure2}a) are higher in females. Mean modal controllability in the subcortex is also higher in females ($r^2 = 0.047$, $df = 876$, $p < 10^{-4}$; Fig. \ref{fig:figure2}b), whereas mean average controllability in the subcortex is higher in males ($r^2 = 0.036$, $df = 876$, $p = 0.034$; Fig. \ref{fig:figure2}b). Only with average controllability in the subcortex did males have higher controllability than females; this result suggests that the connectivity profile of subcortical regions may contribute to sex differences in functional brain dynamics.

In a finer-grained analysis, we investigated whether the controllability of single regions differed by sex. Modal and average controllability were separately examined at each region, while accounting for age, total brain volume, handedness, and mean in-scanner head motion as model covariates. We found that average and modal controllability differed by sex at a subset of network nodes after FDR correction ($q < 0.05$) for multiple comparisons. Nodes with average controllability values that differed by sex were almost all (4/5) higher in males and located in the frontal lobe or subcortex (Fig. \ref{fig:figure2}c-d). Conversely, nodes with modal controllability values that differed by sex were all (18/18) higher in females and located in frontoparietal and subcortical systems. These results suggest that controller strength differs between males and females on a region-specific and control strategy-specific basis.

\subsection*{Development of network controllability across the sexes}

\begin{figure}[h!]
\begin{center}
\includegraphics[width=8.6cm,keepaspectratio]{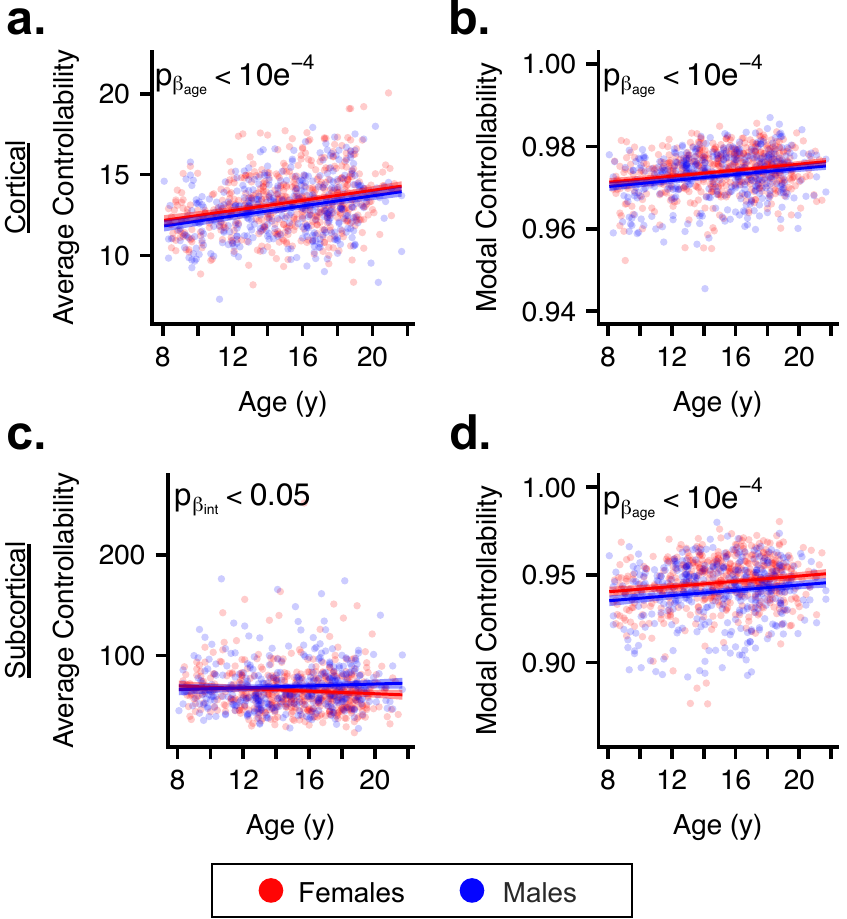}
\caption{{\textbf{Network controllability as a function of age.} \emph{(a-d)} Relationship between age and controllability estimates, averaged over 219 cortical nodes \emph{(a-b)} and 14 subcortical nodes \emph{(c-d)}, for average controllability \emph{(a, c)} and modal controllability \emph{(b, d)}. Modal controllability increases with age in the cortex and in the subcortex; average controllability increases with age in the cortex. In the subcortex, there was a significant age-by-sex interaction for average controllability. However, the simple slopes for males ($\beta_{age}$) and females ($\beta_{age} + \beta_{age\times sex}$) were not significantly different from 0. Age-by-sex interactions in \emph{a, b}, and \emph{d} were not significant. Red and blue represent females and males, respectively. Each point represents a subject's partial residual with respect to age; lines represent regression slopes for males and females separately, and shaded envelopes denote 95\% confidence intervals of the slopes. The $p$ values for $\beta_{age}$ are shown in panels \emph{(a,b,d)} and for $\beta_{age\times sex}$ are shown in panel \emph{(c)}.
\label{fig:figure3}%
}}
\end{center}
\end{figure}

We next turned to assessing whether the developmental arc of network controllability differed by sex. We found that controllability in the cortex and subcortex tended to increase with age, with the exception of average controllability in the subcortex. In the cortex ($r^2 = 0.13$, $df = 876$, $p = 9.9\times10^{-10}$; Fig. \ref{fig:figure3}b) and subcortex ($r^2 = 0.047$, $df = 876$, $p = 1.0\times10^{-5}$; Fig. \ref{fig:figure3}d), mean modal controllability increases with age for males and females. Mean cortical average controllability also increases with age ($r^2 = 0.082$, $df = 876$, $p = 1.9\times10^{-15}$; Fig. \ref{fig:figure3}b). Interestingly, there was a significant age-by-sex interaction only with subcortical average controllability, such that the slope was positive for males and negative for females (Fig. \ref{fig:figure3}c). However, we also found that subcortical average controllability remains stable throughout development for both males ($\beta_{age} = 0.47$, $t = 1.29$, $p = 0.20$, $df = 875$) and females ($(\beta_{age} + \beta_{age*sex}) = -0.65$, $p = 0.052$, $t = -1.94$, $df = 875$) \citep{preacher2004simple}. Taken together, these results suggest that controllability changes with age similarly for males and females, but that average controllability in the subcortex changes with age differently than modal controllability or than cortical average controllability.

\begin{figure}[h!]
\begin{center}
\includegraphics[width=18cm,keepaspectratio]{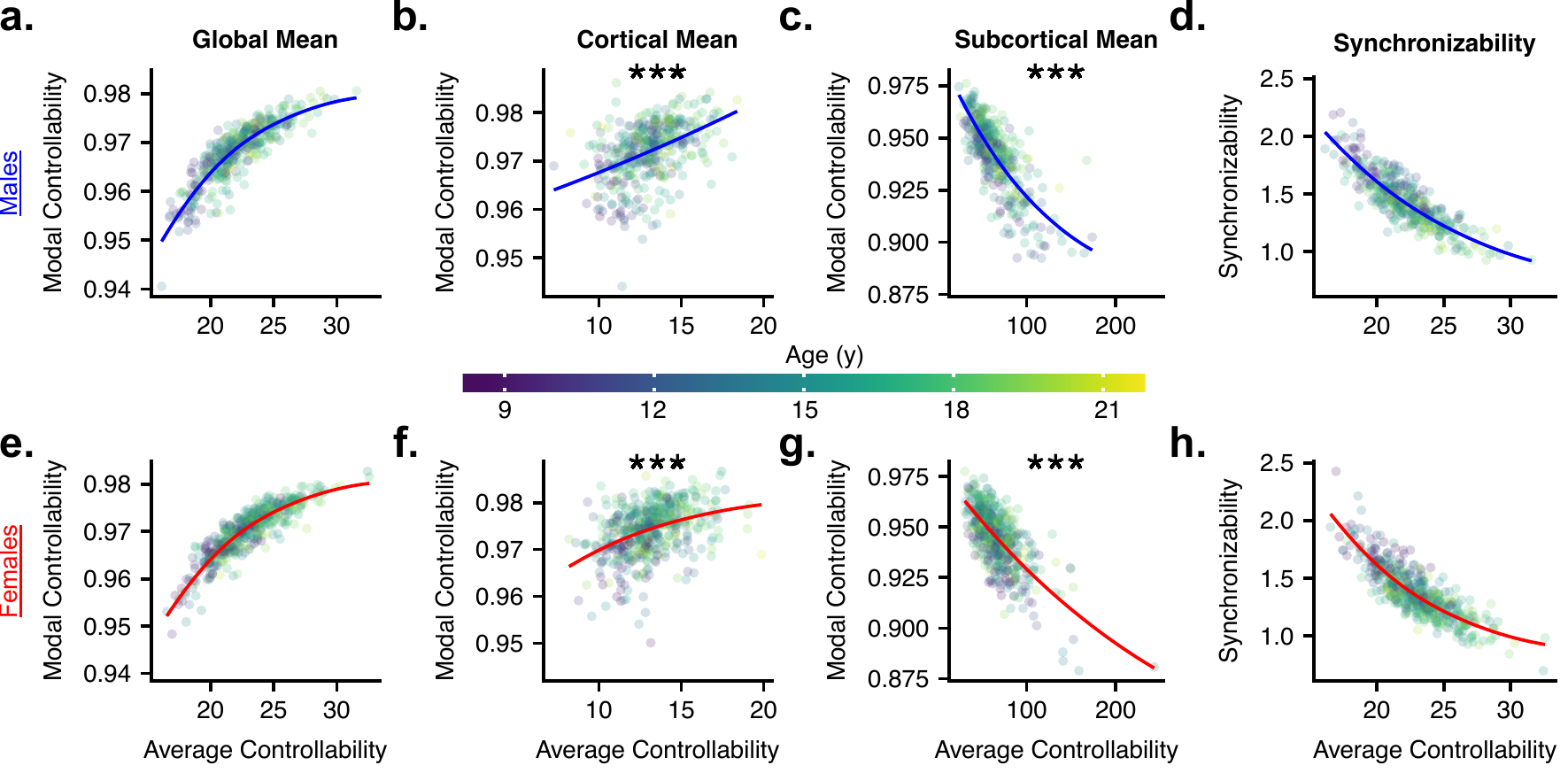}
\caption{{\textbf{Sex effects on relationships between metrics of network control and dynamics.} Relationships between whole-brain average and modal controllability \emph{(a, e)}, between synchronizability and whole-brain average controllability \emph{(d, h)}, and between regional average controllability and regional modal controllability \emph{(b-c, f-g)}. Data from male subjects is depicted in panels \emph{a-d} and data from female subjects is depicted in panels \emph{e-h}. Non-linear least squares was used on the base model: $y = a + b*\exp(c*x)$, and the full model: $y = (a + \alpha*$Sex$) + (b + \beta*$Sex$)*\exp((c+\gamma*$Sex$)*x)$. Slopes represent coefficients from the full model. The $p$ values were obtained from ANOVA comparing base and full models. In these cases, a low $p$ value indicates that allowing for a sex term in the model significantly increases the explained variance. *, $p < 0.05$, **, $p < 0.01$, ***, $p < 10^{-4}$.
\label{fig:figure4}%
}}
\end{center}
\end{figure}

With the knowledge that controllability has a sex-independent relationship with age, we were interested in testing the hypothesis that sex influences the relationship between different types of controllability in the developing brain. Following ref.\citep{tang2017developmental}, we fit the relationships between average and modal controllability with the exponential function $y = a + b\exp(cx)$ separately for each sex (Eq. \ref{eq:equation6}). Our results showed that males and females did not have a statistically different relationship between average and modal controllability averaged across the whole brain ($p = 0.16$, $df = 3$; Fig. \ref{fig:figure4}a,e). In the cortex alone, average and modal controllability followed an increasing exponential form (Fig. \ref{fig:figure4}b,f), similar to that of the whole brain. In contrast, in the subcortex alone average and modal controllability followed a decreasing exponential form (Fig. \ref{fig:figure4}d,h). When sex was included in the model, fits improved significantly (cortex: $p = 2.1\times10^{-8}$, $df = 3$, Fig. \ref{fig:figure4}b,f; subcortex: $p = 6.3\times10^{-9}$, $df = 3$, Fig. \ref{fig:figure4}c,g), suggesting that these regions may contain sex-dependent differences in structural connections important for controlling brain network state.

Next, we performed a specificity analysis to determine whether our results could be further confirmed by sex differences in a contrasting metric -- synchronizability -- that probed the susceptibility of the network to be constrained within a narrow (rather than broad) range of dynamics. We found that synchronizability did not differ between males and females (\ref{fig:figureS1}b) and decreased with age in a sex-independent fashion (\ref{fig:figureS1}d). Moreover, the exponential relationship between synchronizability and average controllability did not differ by sex ($p = 0.33$, $df = 3$; Fig. \ref{fig:figure4}d,h), confirming the local (Fig. \ref{fig:figure2}c-d) rather than global (Fig \ref{fig:figure1_5}b) differences in metrics of network control and dynamics.

\subsection*{Sex differences in network controllability predict individual differences in executive function}

\begin{figure}[h!]
\begin{center}
\includegraphics[width=18cm,keepaspectratio]{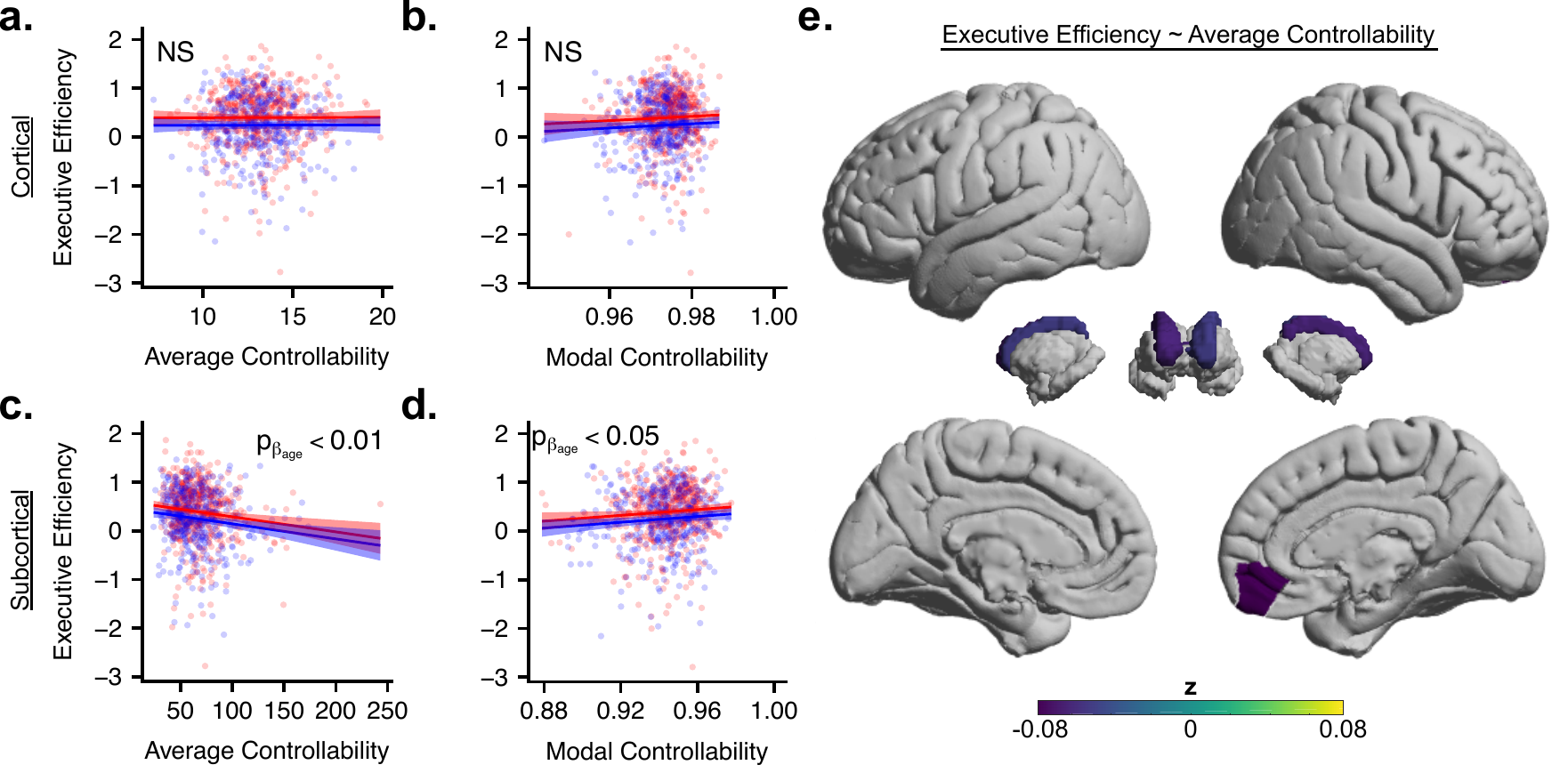}
\caption{{\textbf{Network controllability and cognitive performance.} \emph{(a-d)} Relationship between age and controllability estimates, averaged over 219 cortical nodes \emph{(a-b)} and 14 subcortical nodes \emph{(c-d)}, for average controllability \emph{(a, c)} and modal controllability \emph{(b, d)}. Cortical controllability \emph{(a-b)} is not significantly related to executive function, but subcortical average controllability \emph{(c)} is negatively correlated with executive function while subcortical modal controllability \emph{(d)} is positively correlated with executive function. In panels \emph{a-d}, red and blue represent females and males, respectively. Each data point represents a subject's partial residual with respect to controllability; lines represent regression slopes for males and females separately, and shaded envelopes denote 95\% confidence intervals of the slopes. \emph{(e)} Brain regions with sex-associated average controllability and their relation to executive function (FDR corrected, $q < 0.05$). Colors reflect values of standardized regression coefficients for average controllability at each node as a predictor of executive function while controlling for covariates. Modal controllability was not significantly associated with executive function. For panels \emph{a-e}, there were no significant sex-controllability interactions on executive function. 
\label{fig:figure5}}%
}
\end{center}
\end{figure}

While sex differences in network controllability are of interest in understanding the structural drivers of brain dynamics, their impact on behavior requires a link to cognition. Here we examine the relation between executive function and network controllability across cortex and subcortex separately, and then at individual brain regions. When considering mean cortical and subcortical controllability and executive function we found that neither average ($r^2 = 0.46$, $df = 873$, $p = 0.96$; Fig. \ref{fig:figure5}a) nor modal controllability ($r^2 = 0.46$, $df = 873$, $p = 0.25$; Fig. \ref{fig:figure5}b) in the cortex were associated with performance. A different trend was apparent in the subcortex: average controllability was negatively correlated with performance ($r^2 = 0.47$, $df = 873$, $p = 5.4\times10^{-4}$; Fig. \ref{fig:figure5}c), while modal controllability was positively correlated with performance ($r^2 = 0.47$, $df = 873$, $p = 0.024$; Fig. \ref{fig:figure5}d). 

Next, we considered the 21 brain areas that we had previously found to display sex differences in controllability values (5 nodes for average controllability and 18 nodes for modal controllability, with 2 nodes overlapping). Nodes with significantly higher average controllability in males (right medial orbitofrontal cortex and bilateral caudate nuclei) had average controllability values that were negatively related to executive function (Fig. \ref{fig:figure5}e; FDR corrected, $q < 0.05$). Interestingly, modal controllability at any individual node was not significantly associated with executive function after FDR correction; however, two nodes with uncorrected $p < 0.05$ (right caudate nucleus and left inferior parietal lobe) had modal controllability values that were higher in females and positively associated with executive function. This result suggests that modal controllability of one particular node is not as strong of a predictor of cognitive performance as mean modal controllability over the subcortex as a whole.

\begin{figure}[h!]
	\includegraphics[width=8.6cm,keepaspectratio]{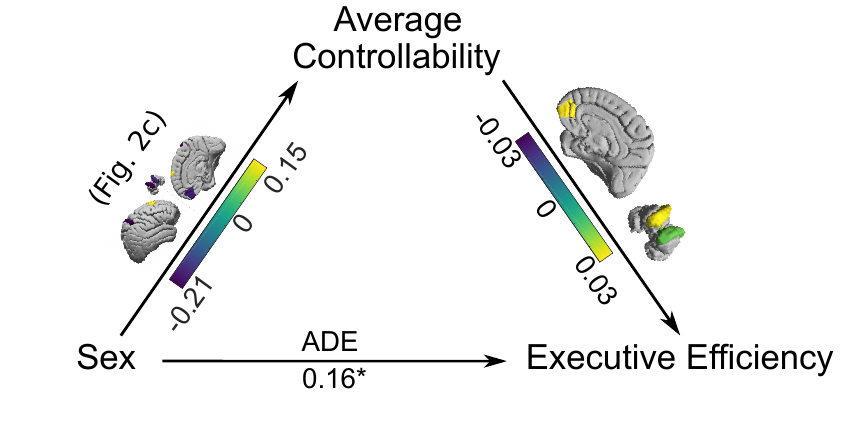}
	\centering
	\begin{minipage}{\linewidth}
		\begin{center}
			\title{\textit{Average Controllability}}
		
		\begin{tabular}{rrrrrrrrr}
			\hline
			Control Node & ACME & $p$ & ADE & $p$\\ 
			\hline
			R Medial orbito frontal, parcel 1 & 0.03 & 0.00 & 0.13 & 0.01 \\ 
			R Superior frontal, parcel 4 & 0.01 & 0.43 & 0.15 & 0.01 \\ 
			R Postcentral, parcel 5 & 0.01 & 0.10 & 0.15 & 0.00 \\ 
			R Caudate & 0.03 & 0.01 & 0.13 & 0.01 \\ 
			L Caudate & 0.02 & 0.03 & 0.14 & 0.01 \\ 
			\hline
		\end{tabular}
	\end{center}
	\end{minipage}
	\bigskip
	
	\caption{{\textbf{Average controllability mediates the relation between sex and executive function.} \emph{(a)} Non-parametric causal mediation analysis for sex $\rightarrow$ average controllability $\rightarrow$ executive function, using control nodes with sex-associated average controllability values. ACME = average causal mediation effect, i.e. average controllability $\rightarrow$ executive function; ADE = average direct effect, i.e. sex $\rightarrow$ executive function. Associations between sex and controllability from Fig. \ref{fig:figure2}c are shown; cooler colors indicate higher controllability in males. 1000 simulations were performed. $p$-values were FDR corrected ($q < 0.05$). Age, total brain volume, handedness, and head motion were included in all analyses. * indicates standardized regression coefficient for sex as a predictor of executive function with the same covariates in the model.
			\label{fig:figure6}%
	}}
	
\end{figure}

One parsimonious explanation for the results reported up to this point is that controllability mediates the relationship between sex and executive function: specifically, subcortical average controllability is higher in males, and increasing subcortical average controllability is associated with poorer performance on an assessment of executive function. We explicitly tested for such a mediation and found that, indeed, average controllability at the right medial orbitofrontal cortex, right superior frontal cortex, right postcentral gyrus and the bilateral caudate nuclei were statistically significant mediators of the relationship between sex and executive function (Fig. \ref{fig:figure6}a). Similarly, subcortical modal controllability was higher in females, and increasing subcortical modal controllability was associated with better executive function. However, consistent with the fact that no individual node had modal controllability values associated with executive function, modal controllability was not found to be a mediator of the relationship between sex and executive function. Taken together, these results suggest that sex differences in average controllability may be a predictive biomarker for sex differences in executive function.

\subsection*{Sex-dependent relationships between n-back task activation magnitudes and controllability}

Describing the relationship between sex, regional controllability, and cognitive performance provides us with a better understanding of the importance of structural brain networks in sex differences in executive function. The final aspect of our hypothesis pertains to whether network control theory can be used to explain how differences in brain network structure produce divergent patterns of brain activity that underpin executive function. We hypothesized that regional controllability values would predict regional n-back task activation magnitudes, and that associations between controllability and activation would differ by sex.

\begin{figure}[h!]
	\begin{center}
		\includegraphics[width=18cm,keepaspectratio]{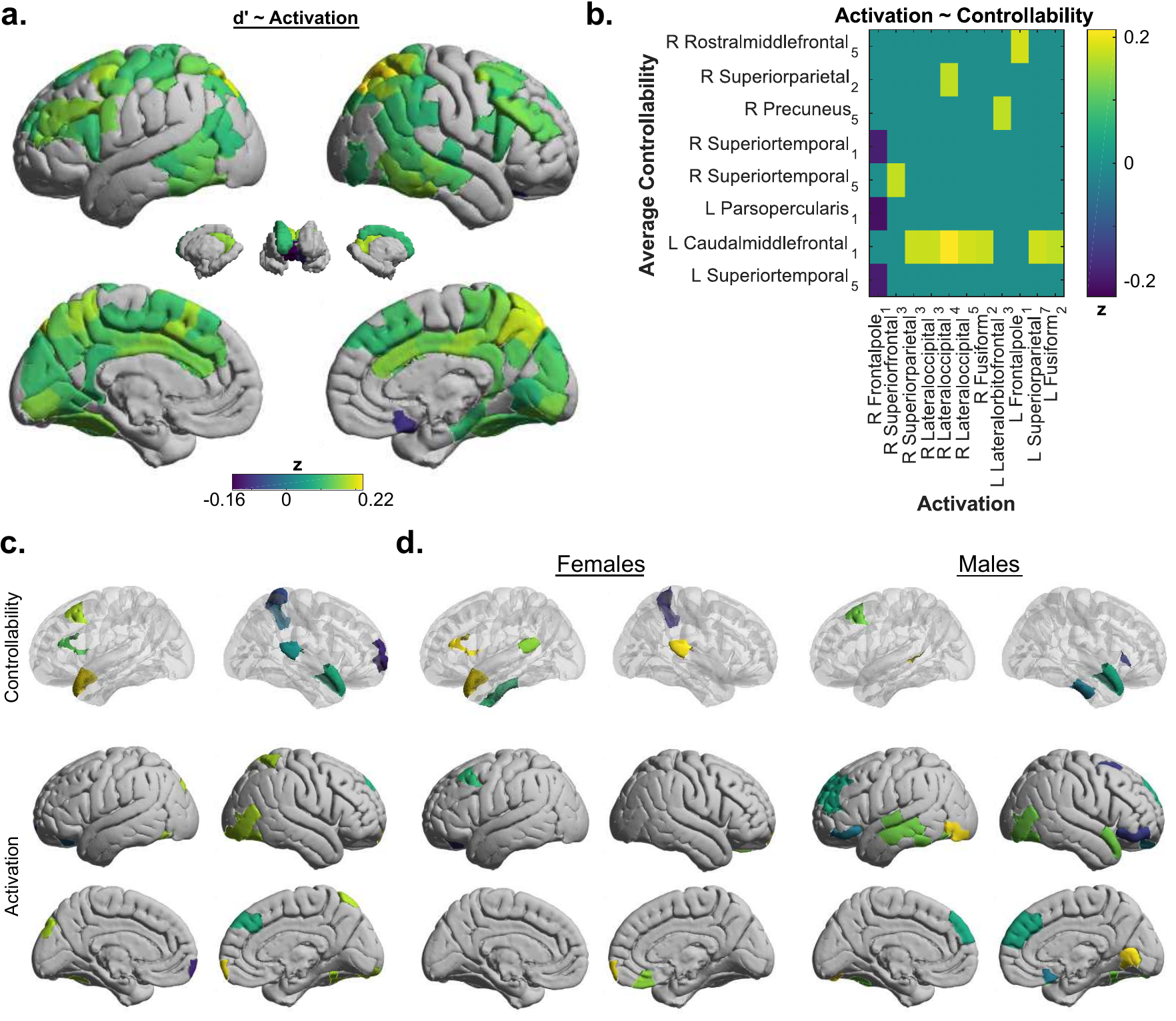}
		\caption{{\textbf{Regional controllability predicts n-back task activation and cognitive performance differently for males and females.} \emph{(a)} Brain regions at which the 2-back minus 0-back GLM $\beta$ weights are associated with 2-back $d^\prime$, a summary measure of task performance in $n = 659$ subjects with quality fMRI data. \emph{(b)} Heatmap depicting standardized multiple linear regression $\beta$'s for regional average controllability as a predictor of regional 2-back minus 0-back activation. Only nodes with associations surviving FDR correction are shown ($q < 0.05$). \emph{(c)} Visual representation of the data presented in panel \emph{(b)}; average controllability at highlighted nodes is associated with activation at nodes of the same colors. \emph{(d)} Associations between activation and average controllability for $n = 283$ males and $n = 376$ females, separately. Modal controllability was not significantly associated with activation. For panels \emph{(c,d)}, corresponding colors indicate association between control node and regional activation.
				\label{fig:figure7}%
		}}
	\end{center}
\end{figure}

\begin{figure}[h!]
	\begin{center}
		\includegraphics[width=18cm,keepaspectratio]{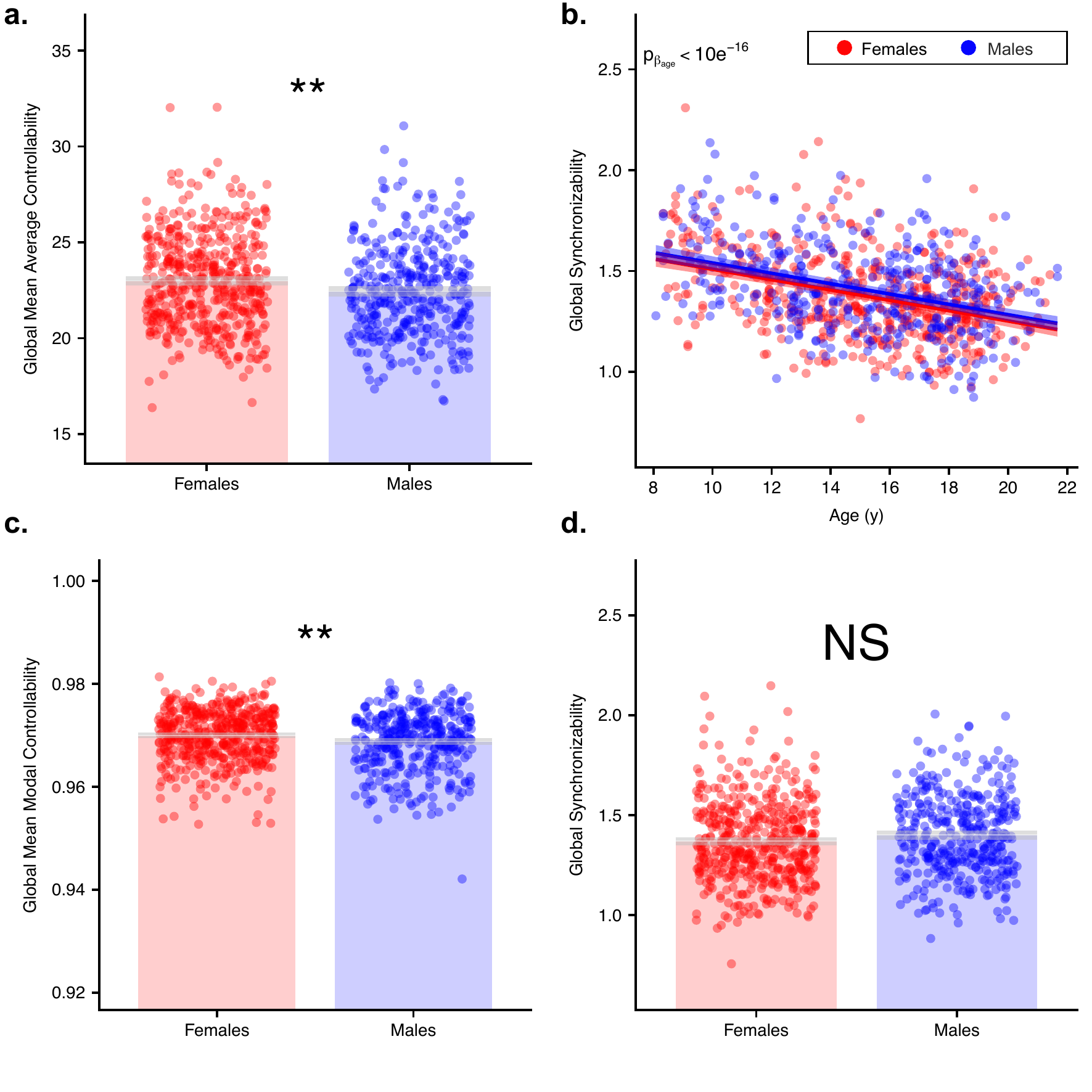}
		\caption{{\textbf{Whole-brain network controllability and synchronizability by sex.} \emph{(a, c)} Average \emph{(a)} and modal \emph{(c)} controllability, averaged across all brain regions (hereafter ``global''), comparing males and females. \emph{(b)} Whole-brain (``global'') synchronizability over age. \emph{(d)} Global synchronizability by sex. We observed no significant age-by-sex interactions on synchronizability. Red and blue represent females and males, respectively. Each point represents a subject's partial residual with respect to sex; bar height represents the predicted mean for males and females separately, and shaded envelopes denote 95\% confidence intervals of the prediction. *, $p < 0.05$, **, $p < 0.01$, ***, $p < 10^{-4}$.
				\label{fig:figureS1}%
		}}
	\end{center}
\end{figure}

To obtain a reference point for activation profiles associated with strong executive function, we separately regressed activation at each node on $d^\prime$ \citep{satterthwaite2013functional}. This analysis identified 113 nodes for which activation was associated with successful task performance (Fig. \ref{fig:figure7}a). Note that in contrast to the use of a general factor score for executive efficiency in the previous analyses, here we considered the $d^\prime$ measure to summarize performance on the n-back task only \citep{snodgrass1988pragmatics} so that the performance measure is directly related to the context in which activation values are measured. Interestingly, sex was not associated with activation at any brain region (FDR corrected, $q < 0.05$, data not shown). 

After identifying regions at which activation was associated with executive function, we sought to relate these activation values to controllability metrics of structural brain networks. We found a complex pattern of both positive and negative relationships between average controllability at 8 nodes with activation at 11 nodes (Fig. \ref{fig:figure7}b). Average controllability at the left caudal middle frontal lobe was associated with activation at 7 different regions, which cluster together in a symmetric fashion in superior parietal and lateral occipital cortex (Fig. \ref{fig:figure7}c). Modal controllability was not significantly associated with activation. 

Next, we tested our hypothesis that the relationship between regional controllability and activation depends on sex. Specifically, we performed the same analysis on males and females, separately. Consistent with the pooled analysis, the profile of controllability-activation associations overlapped significantly between the sexes (Fig. \ref{fig:figure7}d, Fig. \ref{fig:figureS4}a). While there was no overlap in the precise regions involved in these sex-split analyses, there were gross anatomical similarities between males and females. In both sexes, controllability at regions near the temporal poles was associated with activation at medial frontal regions. We observed no significant interaction between controllability and sex in predicting activation (Fig. \ref{fig:figureS4}b), although the directionality of the interaction terms was consistent with the observed sex differences (Fig. \ref{fig:figureS4}a-b). These results support the notion that both similarities and differences in structure-function relationships exist across the sexes.

\subsection*{Weighted degree does not explain controllability trends}

\begin{figure}[h!]
	\begin{center}
		\includegraphics[width=18cm,keepaspectratio]{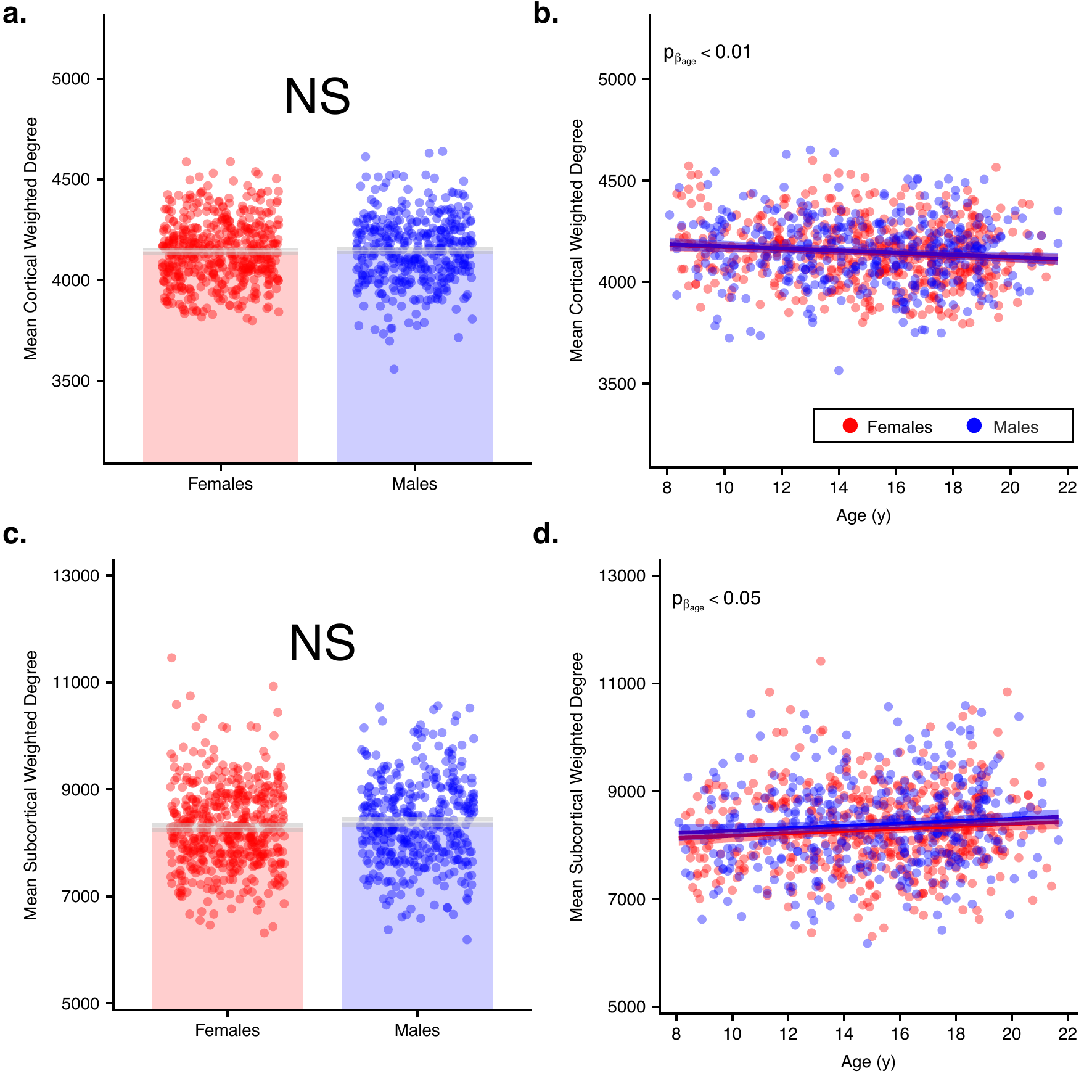}
		\caption{{\textbf{Trends between regional weighted degree, age, and sex differ from those with network controllability.} \emph{(a-d)} Weighted degree averaged over 219 cortical nodes \emph{(a-b)} and 14 subcortical nodes \emph{(c-d)}. Panels \emph{(a, c)} show that no significant sex differences exist in mean cortical \emph{(a)} or subcortical \emph{(c)} weighted degree. Panels \emph{(b, d)} show that weighted degree decreases with age in the cortex \emph{(b)} and increases with age in the subcortex \emph{(d)}. We observed no significant age-by-sex interactions in these models. Red and blue represent females and males, respectively. Each point represents a subject's partial residual with respect to sex; bar height represents the predicted mean for males and females separately, and shaded envelopes denote 95\% confidence intervals of the prediction. *, $p < 0.05$, **, $p < 0.01$, ***, $p < 10^{-4}$.
				\label{fig:figureS2}%
		}}
	\end{center}
\end{figure}

\begin{figure}[h!]
	\begin{center}
		\includegraphics[width=8.6cm,keepaspectratio]{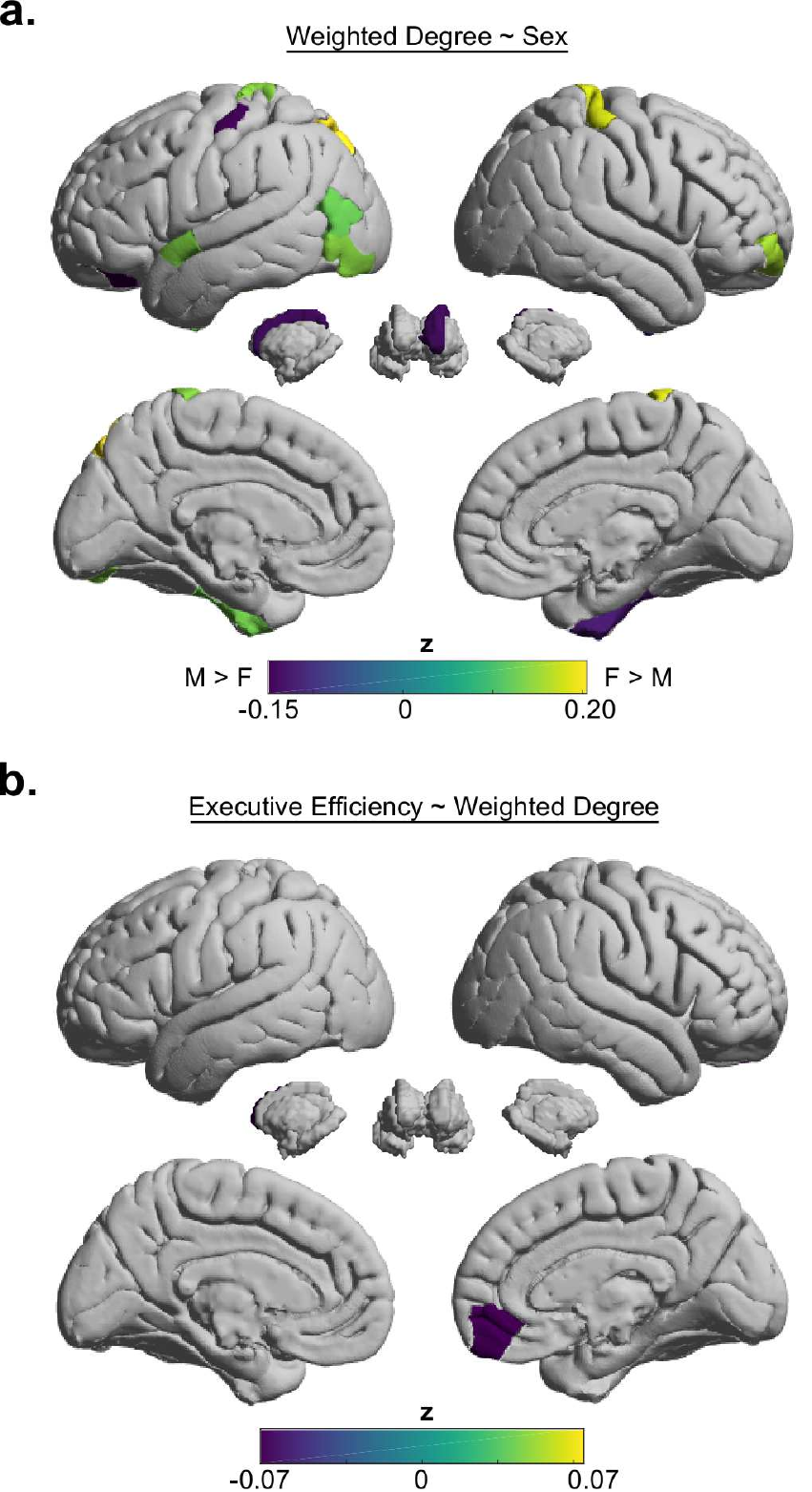}
		\caption{{\textbf{Relationship between sex, network controllability, and cognitive performance is not well-explained by weighted degree.} \emph{(a)} Heatmaps where color intensity corresponds to standardized regression coefficient for sex, with warmer colors indicating higher weighted degree in females. Five out of twenty-one regions with sex-associated controllability also have sex-associated weighted degree. Of those five, only average controllability values at the left caudate nucleus and right postcentral gyrus have the same direction of assocation with sex as weighted degree. \emph{(b)} Association between executive function and weighted degree at nodes with sex-associated average controllability, modal controllability, or weighted degree, correcting for multiple comparisons with an FDR correction at $q < 0.05$. No sex $\times$ weighted degree interaction terms on executive function were significant. Age, total brain volume, handedness, and head motion were included as covariates in all analyses.
				\label{fig:figureS3}%
		}}
	\end{center}
\end{figure}

\begin{figure}[h!]
	\begin{center}
		\includegraphics[width=18cm,keepaspectratio]{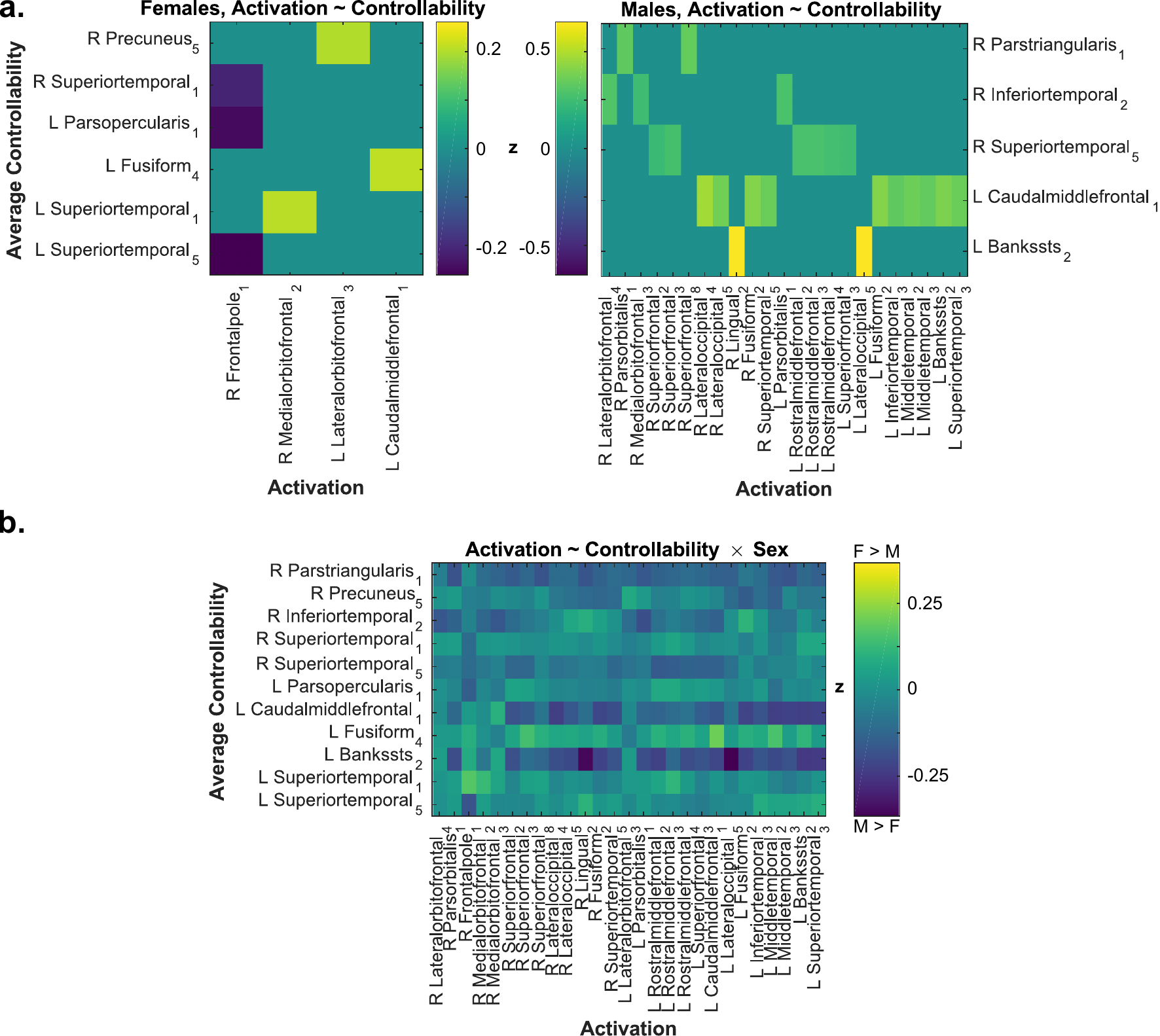}
		\caption{{\textbf{Sex-dependent relationships between network controllability and activation.} \emph{(a)} Heatmap of standardized multiple linear regression $\beta$'s for regional average controllability as a predictor of regional 2-back minus 0-back activation. \emph{(Left)} Data from $n = 376$ females. \emph{(Right)} Data from $n = 283$ males.  Only nodes with associations surviving FDR correction are shown ($q < 0.05$). \emph{(b)} Heatmap of standardized $\beta$'s for average controllability $\times$ sex interaction. \textit{No $\beta$ was significant after FDR correction ($q < 0.05$)}, but nodes from panel \emph{(a)} are shown to illustrate consistent directionality of association. Especially obvious are the left caudal middle frontal (parcel 1) and left banks of STS (parcel 2), where darker colors on the interaction heatmap indicate a larger effect size in males. The left fusiform gyrus (parcel 4) shows the opposite trend, with lighter colors on the interaction heatmap indicating a larger effect size in females.
				\label{fig:figureS4}%
		}}
	\end{center}
\end{figure}

Weighted degree has previously been shown to correlate well with controllability metrics \citep{gu2015controllability}. We performed several specificity analyses to determine whether our results could be trivially explained by sex-differences in the weighted degree of nodes in the network. We identified 13 nodes where weighted degree was associated with sex (Fig. \ref{fig:figureS3}a), and found that only 5/21 unique nodes with sex-associated controllability values also had a significant association between weighted degree and sex. Among those 5 unique nodes, only average controllability at the left caudate nucleus and the right post central gyrus had the same directionality of association with sex as the association with weighted degree (i.e. higher in males and females, respectively). When weighted degree was averaged across the anatomical cortex or subcortex, there were no significant associations between sex and cortical ($r^2 = 0.090$, $df = 876$, $p = 0.73$; Fig. \ref{fig:figureS2}a) or subcortical ($r^2 = 0.077$, $df = 876$, $p = 0.13$; Fig. \ref{fig:figureS2}c) weighted degree. These findings suggest that despite the known associations between controllability and weighted degree \citep{gu2015controllability}, weighted degree does not explain the observed sex differences in controllability.

Next, we tested whether developmental trends of weighted degree paralleled those of controllability metrics. We found that when averaged across the cortex or subcortex, weighted degree is positively associated with average controllability and negatively associated with modal controllability. However, mean weighted degree increased with age in the cortex ($r^2 = 0.090$, $df = 876$, $p = 0.0034$; Fig. \ref{fig:figureS2}b), but decreased with age in the subcortex ($r^2 = 0.077$, $df = 876$, $p = 0.011$; Fig. \ref{fig:figureS2}d). Neither of these two trends (Fig. \ref{fig:figureS2}b,d) were simple mirrors of the relationships between age and regional average or modal controllability (Fig. \ref{fig:figure3}), suggesting that developmental trends of controllability reflect a more complex phenomenon than developmental trends of weighted degree. Furthermore, at nodes with sex-associated weighted degree or sex-associated controllability, weighted degree did not significantly mediate the relationship between sex and executive function (Fig. \ref{fig:figureS6}). These findings suggest that weighted degree does not explain developmental trends of controllability, nor does it mediate the relationship between sex and executive function.

\begin{figure}[h!]
	\begin{center}
		\includegraphics[width=18cm,keepaspectratio]{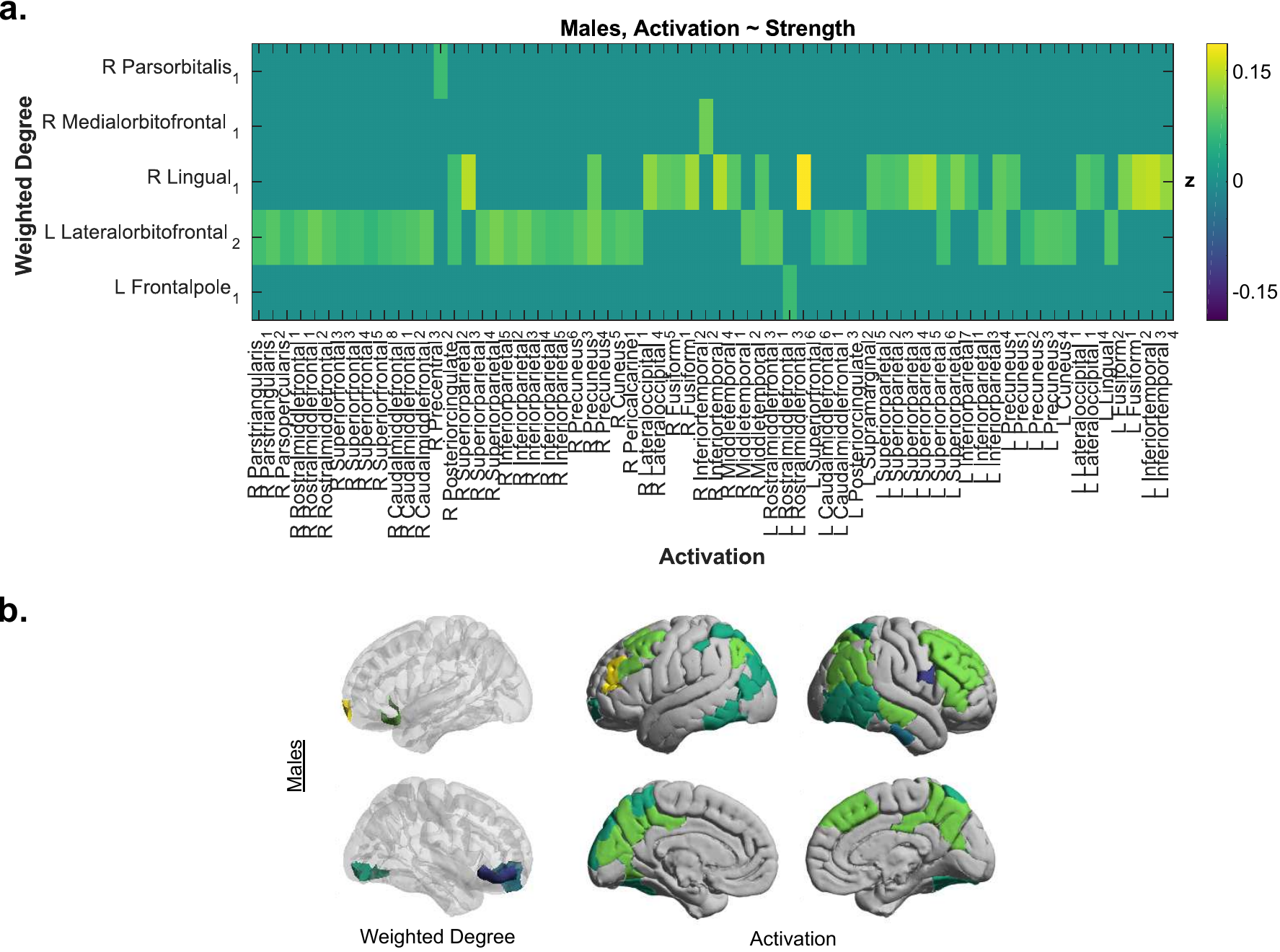}
		\caption{{\textbf{Relationship between weighted degree and activation.} \emph{(a)} Heatmap of multiple linear regression $\beta$'s for regional weighted degree as a predictor of regional 2-back minus 0-back activation. Only nodes with associations surviving FDR correction are shown ($q < 0.05$). There were only significant associations in the sample of $n = 283$ males, and not in the pooled sample or in females alone. \emph{(b)} Visualization of associations between weighted degree and activation. Weighted degree at highlighted nodes is associated with activation at nodes of the same colors; that is, corresponding colors indicate the association between structural connections and regional activation.
				\label{fig:figureS5}%
		}}
	\end{center}
\end{figure}

Finally, we assessed whether the associations between controllability and activation could be simply explained by the weighted degree of nodes in the network. We found no significant associations between activation and weighted degree in a pooled analysis. Moreover, including each subject's global mean weighted degree in the model did not change the associations between average controllability and activation (data not shown). While there were significant associations between activation and weighted degree for males alone (Fig. \ref*{fig:figureS5}), weighted degree-activation profiles did not overlap well with average controllability-activation profiles. These results suggest that weighted degree does not explain the relation between controllability and activation. 

\section*{Discussion}

Our study demonstrates that network control theory can be used to explain how differences in brain network structure produce divergent patterns of brain activity that underpin executive function and its differences across males and females. We first showed that the relative locations of controllers by strength did not differ between males and females, suggesting that the overall structure of control points is similar across sex. Then, we showed that controllability covaried with age in a sex-independent fashion in both the cortex and subcortex. While global and developmental sex differences appear minimal, local sex differences in controllability exist and average controllability at those regions predicts executive function. Notably, average controllability significantly mediated the association between sex and executive function, such that subcortical average controllability was higher in males and is negatively associated with executive function. Crucially, consistent with predictions from network control theory, we also observed sex-dependent associations between nodal average controllability and n-back task BOLD activation magnitudes, demonstrating that differences in brain network structure produce divergent patterns of brain activity supporting executive function. 

\subsection*{Implications for cognitive and clinical neuroscience}

Executive function is generally lower in males, which is reflected in higher rates of impulsivity \citep{chap}, ADHD diagnoses \citep{willcutt2012prevalence}, criminality \citep{cross2011sex}, and substance use \citep{ROMER20092916}. While it is known that sex differences exist in the prevalence of disorders of executive function and the putative neural circuitry involved \citep{castellanos2002neuroscience,pohjalainen1998sex}, it is unclear whether the pathophysiology of such disorders is sex-specific or sex-independent. Our study significantly extends the boundaries of knowledge in demonstrating neurophysiological markers of sex differences in executive function, and in couching such markers within a general network control theory of brain function.

The present work also provides important groundwork for clinical therapy. The delivery of psychiatric care is shifting towards personalized, targeted interventions. This shift can be supported by the methodology and computational tools of network neuroscience, where accurate models of brain structure and dynamics can be constructed for single individuals \citep{STEPHAN2017180,izhikevich2008large}. Such subject-, age-, and sex-specific models of brain structure and function can directly inform neuromodulatory therapies such as transcranial magnetic stimulation, by offering predictions about how stimulation will affect both the area being stimulated, and other areas connected to it, thereby producing a complex spatiotemporal influence on brain state. Specifically, assuming a model of brain dynamics allows us to determine which nodes could most easily drive transitions in the state of brain activity via some stimulatory input. The network control theory that we use here to study the internal modulation of brain state (via undertaking a task requiring executive function) also makes explicit predictions about the external modulation of brain state (via stimulation or neurofeedback) \cite{murphy2017network}. Indeed, average controllability values were associated with sex-dependent, symmetric patterns of brain activity (Fig. \ref{fig:figure7}c-d). Our results suggest that average controllability values contain useful information about brain state, beyond what can be predicted from a simple streamline-weighted adjacency matrix. These observations are the first steps towards a characterization of brain dynamics that would allow clinicians to predict the impact of stimulation given a subject-, age-, and sex-specific network architecture, thereby producing predictable changes in brain state.

An understanding of the relation between network controllability, sex, and executive function could provide important context for the study and diagnosis of neurological disease and psychiatric disorders whose prevalence may differ by sex and whose presentation includes alterations in executive function. In our study, the most robust associations between controllability, sex, and executive function are found in the subcortex and the prefrontal cortex. It is widely known that the prefrontal cortex is important for behavioral planning and working memory \citep{tanji2008role,euston2012role}, two key components of executive function. The subcortex alone is less commonly associated with executive function, but there is a strong precedent for the requirement of circuitry between dorsolateral prefrontal cortex (DLPFC) and subcortical structures for executive function \citep{cummings1993frontsubc,duke2000executive,bonelli2007frontal,tekin2002frontal}. Vascular \citep{Bombois2595} or neurodegenerative \citep{tekin2002frontal} (i.e. Huntington's disease, Parkinson's disease) lesions to the subcortex give rise to executive dysfunction, similar to what is observed with DLPFC lesions. Moreover, ADHD, a disorder characterized by impulsivity and executive dysfunction, is associated with decreased striatal dopamine transporter expression \citep{castellanos2002neuroscience}, aberrant striatal fMRI BOLD activation \citep{PLICHTA20097}, and decreased striatal volume \citep{CUBILLO2012194,vaidya2011neurodevelopmental}. Sex differences in striatal volume \citep{raz1995age} and striatal dopamine receptor binding \citep{pohjalainen1998sex} have also been reported. In light of these results, our study provides an important account of sex differences in subcortical connectivity from the dynamical perspective of network control theory. Our results suggest that the connectivity profile of subcortical structures is related to sex and is important for executive function, supporting the notion that disorders of executive function arise in part from altered structural connectivity.

\subsection*{Methodological Considerations}

Several methodological considerations are pertinent to this work. First, we use a time-invariant, linear model of brain dynamics because its network control properties have been well characterized mathematically; however, it is known that the brain is highly non-linear \citep{FREEMAN1985147}. Yet, the associations between controllability and brain activity that we uncover here suggest that a simple linear model is sufficient to capture some aspect of the underlying brain dynamics. Second, axons transmit information in a unidirectional fashion, but diffusion imaging and associated tractography tools cannot elucidate the directionality of large axonal fiber bundles. Thus, we construct a symmetric adjacency matrix $\mathbf{A}$, assuming bidirectional influence between network nodes, and all interpretations of regional controllability depend on the realism of that assumption. Third, the eigenvalues of a symmetric matrix can only ever be real and thus the system will not oscillate, as neural systems are known to do. Finally, streamline counts and network density vary between scans \citep{zhu2011quantification} and with different tract reconstruction methods \citep{maier2017challenge}, which can complicate interpretation. Nevertheless, numerous DTI-based studies of neuropsychiatric illness have described differences in structural brain networks consistent with clinical and neuroscientific priors \citep{kubicki2007review,pievani2011functional}, suggesting that DTI is capable of capturing subtle but meaningful variation in structural connectivity.

It is also important to acknowledge that our analysis of sex differences focuses on sex assigned at birth, which we refer to as ``sex" throughout this paper. Our data does not include endocrinological measurements relevant to sex, nor does it include any psychosocial assessment of gender identity. Thus, we were not able to control for or assess gender-based differences in brain structure. Furthermore, while there exist two distinct classes of human genitalia, this fact does not imply that brains are also sexually dimorphic \citep{joel2015sex}. Both ``male" and ``female" features exist in both male and female brains, though some features are more common in one sex than the other \citep{joel2015sex,joel2017incorporating}. As a result, there may be more meaningful variance in neurologic phenotypes within each sex than between sexes. The divergent controllability-activation profiles in Fig. \ref{fig:figure7}d may be in part due to this fact. Nevertheless, biological sex is an easily measured variable and identifying correlates of sex makes it a useful biomarker. In the future, identification of additional covariates might help uncover a more ubiquitous reason for sex-associated brain features \citep{joel2017incorporating}.

\section*{Conclusion and Future Directions}

First and foremost, our analysis of sex differences in structural brain networks showed that males and females are highly similar from the perspective of network controllability. The organization of relative controller strength was almost identical between males and females, and sex was not significantly associated with controllability values at most brain regions. However, the differences in average controllability between males and females predicted differences in cognitive performance and effects were most robust in subcortical and frontal regions. Given that BOLD signal is associated with network controllability, an interesting future study might use time-invariant, linear dynamics to predict changes in brain activity after stimulation of regions with high versus low controllability. Our results pave the way for a future study to use sex and age as features that may delineate the likely responses to brain stimulation \emph{a priori}, obviating the need for pre-stimulation diffusion image acquisition and processing. Such an approach would allow the delivery of personalized psychiatric care at low cost with minimal requirements for imaging equipment and computational power.

\section*{Acknowledgments}

This work was supported by an administrative supplement to NIH R21-M MH-106799 (Satterthwaite/Bassett MPI). DSB also acknowledges support from the John D. and Catherine T. MacArthur Foundation, the Alfred P. Sloan Foundation, the Army Research Office through contract number W911NF-14-1-0679, the Army Research Laboratory through contract number W911NF-10-2-0022, the National Institute of Health (2-R01-DC-009209-11, 1R01HD086888-01, R01-MH107235, R01-MH107703, R01MH109520, 1R01NS099348 and R21-M MH-106799), the Office of Naval Research, and the National Science Foundation (BCS-1441502, CAREER PHY-1554488, BCS-1631550, and CNS-1626008). TDS was supported by R01MH107703. DRR was supported by K01MH102609. FP acknowledges support from NSF-BCS-1631112 and NSF-BCS-1430279. The content is solely the responsibility of the authors and does not necessarily represent the official views of any of the funding agencies.

\newpage
\bibliographystyle{unsrtnat}
\bibliography{control_sync_DB.bib}

\begin{thebibliography}{80}
\providecommand{\natexlab}[1]{#1}
\providecommand{\url}[1]{\texttt{#1}}
\expandafter\ifx\csname urlstyle\endcsname\relax
  \providecommand{\doi}[1]{doi: #1}\else
  \providecommand{\doi}{doi: \begingroup \urlstyle{rm}\Url}\fi

\bibitem[Anderson et~al.(2001)Anderson, Anderson, Northam, Jacobs, and
  Catroppa]{anderson2001development}
Vicki~A Anderson, Peter Anderson, Elisabeth Northam, Rani Jacobs, and Cathy
  Catroppa.
\newblock Development of executive functions through late childhood and
  adolescence in an australian sample.
\newblock \emph{Developmental neuropsychology}, 20\penalty0 (1):\penalty0
  385--406, 2001.

\bibitem[Romer et~al.(2009)Romer, Betancourt, Giannetta, Brodsky, Farah, and
  Hurt]{ROMER20092916}
Daniel Romer, Laura Betancourt, Joan~M. Giannetta, Nancy~L. Brodsky, Martha
  Farah, and Hallam Hurt.
\newblock Executive cognitive functions and impulsivity as correlates of risk
  taking and problem behavior in preadolescents.
\newblock \emph{Neuropsychologia}, 47\penalty0 (13):\penalty0 2916 -- 2926,
  2009.
\newblock ISSN 0028-3932.
\newblock \doi{http://dx.doi.org/10.1016/j.neuropsychologia.2009.06.019}.
\newblock URL
  \url{http://www.sciencedirect.com/science/article/pii/S002839320900270X}.

\bibitem[Barkley et~al.(2002)Barkley, Murphy, Dupaul, and
  Bush]{barkley2002driving}
Russell~A Barkley, Kevin~R Murphy, George~J Dupaul, and Tracie Bush.
\newblock Driving in young adults with attention deficit hyperactivity
  disorder: knowledge, performance, adverse outcomes, and the role of executive
  functioning.
\newblock \emph{Journal of the International Neuropsychological Society},
  8\penalty0 (5):\penalty0 655--672, 2002.

\bibitem[Biederman et~al.(2007)Biederman, Petty, Fried, Doyle, Spencer,
  Seidman, Gross, Poetzl, and Faraone]{bied}
J.~Biederman, C.~R. Petty, R.~Fried, A.~E. Doyle, T.~Spencer, L.~J. Seidman,
  L.~Gross, K.~Poetzl, and S.~V. Faraone.
\newblock Stability of executive function deficits into young adult years: a
  prospective longitudinal follow-up study of grown up males with adhd.
\newblock \emph{Acta Psychiatrica Scandinavica}, 116\penalty0 (2):\penalty0
  129--136, 2007.
\newblock ISSN 1600-0447.
\newblock \doi{10.1111/j.1600-0447.2007.01008.x}.
\newblock URL \url{http://dx.doi.org/10.1111/j.1600-0447.2007.01008.x}.

\bibitem[Chapple and Johnson(2007)]{chap}
Constance~L. Chapple and Katherine~A. Johnson.
\newblock Gender differences in impulsivity.
\newblock \emph{Youth Violence and Juvenile Justice}, 5\penalty0 (3):\penalty0
  221--234, 2007.
\newblock \doi{10.1177/1541204007301286}.
\newblock URL \url{http://dx.doi.org/10.1177/1541204007301286}.

\bibitem[Willcutt(2012)]{willcutt2012prevalence}
Erik~G Willcutt.
\newblock The prevalence of dsm-iv attention-deficit/hyperactivity disorder: a
  meta-analytic review.
\newblock \emph{Neurotherapeutics}, 9\penalty0 (3):\penalty0 490--499, 2012.

\bibitem[Cross et~al.(2011)Cross, Copping, and Campbell]{cross2011sex}
Catharine~P Cross, Lee~T Copping, and Anne Campbell.
\newblock Sex differences in impulsivity: a meta-analysis.
\newblock \emph{Psychological bulletin}, 137\penalty0 (1):\penalty0 97, 2011.

\bibitem[Hosenbocus and Chahal(2012)]{hosenbocus2012review}
Sheik Hosenbocus and Raj Chahal.
\newblock A review of executive function deficits and pharmacological
  management in children and adolescents.
\newblock \emph{Journal of the Canadian Academy of Child and Adolescent
  Psychiatry}, 21\penalty0 (3):\penalty0 223, 2012.

\bibitem[Gogtay et~al.(2004)Gogtay, Giedd, Lusk, Hayashi, Greenstein, Vaituzis,
  Nugent, Herman, Clasen, Toga, Rapoport, and Thompson]{gogtay2004dynamic}
N~Gogtay, J~N Giedd, L~Lusk, K~M Hayashi, D~Greenstein, A~C Vaituzis, T~F~3rd
  Nugent, D~H Herman, L~S Clasen, A~W Toga, J~L Rapoport, and P~M Thompson.
\newblock Dynamic mapping of human cortical development during childhood
  through early adulthood.
\newblock \emph{Proc Natl Acad Sci U S A}, 101\penalty0 (21):\penalty0
  8174--8179, 2004.

\bibitem[Baum et~al.(2017)Baum, Ciric, Roalf, Betzel, Moore, Shinohara, Kahn,
  Vandekar, Rupert, Quarmley, Cook, Elliott, Ruparel, Gur, Gur, Bassett, and
  Satterthwaite]{baum2017modular}
G~L Baum, R~Ciric, D~R Roalf, R~F Betzel, T~M Moore, R~T Shinohara, A~E Kahn,
  S~N Vandekar, P~E Rupert, M~Quarmley, P~A Cook, M~A Elliott, K~Ruparel, R~E
  Gur, R~C Gur, D~S Bassett, and T~D Satterthwaite.
\newblock Modular segregation of structural brain networks supports the
  development of executive function in youth.
\newblock \emph{Curr Biol}, 27\penalty0 (11):\penalty0 1561--1572.e8, 2017.

\bibitem[Schmithorst et~al.(2015)Schmithorst, Vannest, Lee, Hernandez-Garcia,
  Plante, Rajagopal, Holland, and Consortium]{schmithorst2015evidence}
V~J Schmithorst, J~Vannest, G~Lee, L~Hernandez-Garcia, E~Plante, A~Rajagopal,
  S~K Holland, and CMIND~Authorship Consortium.
\newblock Evidence that neurovascular coupling underlying the {BOLD} effect
  increases with age during childhood.
\newblock \emph{Hum Brain Mapp}, 36\penalty0 (1):\penalty0 1--15, 2015.

\bibitem[Nomi et~al.(2017)Nomi, Bolt, Ezie, Uddin, and Heller]{nomi2017moment}
J~S Nomi, T~S Bolt, C~E~C Ezie, L~Q Uddin, and A~S Heller.
\newblock Moment-to-moment {BOLD} signal variability reflects regional changes
  in neural flexibility across the lifespan.
\newblock \emph{J Neurosci}, 37\penalty0 (22):\penalty0 5539--5548, 2017.

\bibitem[Keulers et~al.(2011)Keulers, Stiers, and
  Jolles]{keulers2011developmental}
E~H Keulers, P~Stiers, and J~Jolles.
\newblock Developmental changes between ages 13 and 21 years in the extent and
  magnitude of the {BOLD} response during decision making.
\newblock \emph{Neuroimage}, 54\penalty0 (2):\penalty0 1442--1454, 2011.

\bibitem[Fair et~al.(2009)Fair, Cohen, Power, Dosenbach, Church, Miezin,
  Schlaggar, and Petersen]{fair2009functional}
D~A Fair, A~L Cohen, J~D Power, N~U Dosenbach, J~A Church, F~M Miezin, B~L
  Schlaggar, and S~E Petersen.
\newblock Functional brain networks develop from a ``local to distributed''
  organization.
\newblock \emph{PLoS Comput Biol}, 5\penalty0 (5):\penalty0 e1000381, 2009.

\bibitem[Gennatas et~al.(2017)Gennatas, Avants, Wolf, Satterthwaite, Ruparel,
  Ciric, Hakonarson, Gur, and Gur]{Gennatas3550-16}
Efstathios~D. Gennatas, Brian~B. Avants, Daniel~H. Wolf, Theodore~D.
  Satterthwaite, Kosha Ruparel, Rastko Ciric, Hakon Hakonarson, Raquel~E. Gur,
  and Ruben~C. Gur.
\newblock Age-related effects and sex differences in gray matter density,
  volume, mass, and cortical thickness from childhood to young adulthood.
\newblock \emph{Journal of Neuroscience}, 2017.
\newblock ISSN 0270-6474.
\newblock \doi{10.1523/JNEUROSCI.3550-16.2017}.
\newblock URL
  \url{http://www.jneurosci.org/content/early/2017/04/21/JNEUROSCI.3550-16.2017}.

\bibitem[Blakemore and Choudhury(2006)]{blake}
Sarah-Jayne Blakemore and Suparna Choudhury.
\newblock Development of the adolescent brain: implications for executive
  function and social cognition.
\newblock \emph{Journal of Child Psychology and Psychiatry}, 47\penalty0
  (3-4):\penalty0 296--312, 2006.
\newblock ISSN 1469-7610.
\newblock \doi{10.1111/j.1469-7610.2006.01611.x}.
\newblock URL \url{http://dx.doi.org/10.1111/j.1469-7610.2006.01611.x}.

\bibitem[Ingalhalikar et~al.(2014)Ingalhalikar, Smith, Parker, Satterthwaite,
  Elliott, Ruparel, Hakonarson, Gur, Gur, and Verma]{sattconnsex2014}
Madhura Ingalhalikar, Alex Smith, Drew Parker, Theodore~D. Satterthwaite,
  Mark~A. Elliott, Kosha Ruparel, Hakon Hakonarson, Raquel~E. Gur, Ruben~C.
  Gur, and Ragini Verma.
\newblock Sex differences in the structural connectome of the human brain.
\newblock \emph{Proceedings of the National Academy of Sciences}, 111\penalty0
  (2):\penalty0 823--828, 2014.
\newblock \doi{10.1073/pnas.1316909110}.
\newblock URL \url{http://www.pnas.org/content/111/2/823.abstract}.

\bibitem[Satterthwaite et~al.(2014{\natexlab{a}})Satterthwaite, Wolf, Roalf,
  Ruparel, Erus, Vandekar, Gennatas, Elliott, Smith, Hakonarson,
  et~al.]{satterthwaite2014linked}
Theodore~D Satterthwaite, Daniel~H Wolf, David~R Roalf, Kosha Ruparel, Guray
  Erus, Simon Vandekar, Efstathios~D Gennatas, Mark~A Elliott, Alex Smith,
  Hakon Hakonarson, et~al.
\newblock Linked sex differences in cognition and functional connectivity in
  youth.
\newblock \emph{Cerebral cortex}, 25\penalty0 (9):\penalty0 2383--2394,
  2014{\natexlab{a}}.

\bibitem[Liu et~al.(2011)Liu, Slotine, and Barabasi]{liu2011controllability}
Y~Y Liu, J~J Slotine, and A~L Barabasi.
\newblock Controllability of complex networks.
\newblock \emph{Nature}, 473\penalty0 (7346):\penalty0 167--173, 2011.

\bibitem[Pasqualetti et~al.(2014)Pasqualetti, Zampieri, and
  Bullo]{pasqualetti2014controllability}
Fabio Pasqualetti, Sandro Zampieri, and Francesco Bullo.
\newblock Controllability metrics, limitations and algorithms for complex
  networks.
\newblock \emph{IEEE Transactions on Control of Network Systems}, 1\penalty0
  (1):\penalty0 40--52, 2014.

\bibitem[Kailath(1980)]{TK:80}
T~Kailath.
\newblock \emph{Linear Systems}.
\newblock Prentice Hall, Englewood Cliffs, 1980.

\bibitem[Kalman et~al.(1963)Kalman, Ho, and Narendra]{kalman1963}
R.~E. Kalman, Y.~C. Ho, and K.~S. Narendra.
\newblock Controllability of linear dynamical systems.
\newblock \emph{Contributions to Differential Equations}, 1:\penalty0 189--213,
  1963.

\bibitem[Muldoon et~al.(2016)Muldoon, Pasqualetti, Gu, Cieslak, Grafton,
  Vettel, and Bassett]{muldoon2016stimulation}
S~F Muldoon, F~Pasqualetti, S~Gu, M~Cieslak, S~T Grafton, J~M Vettel, and D~S
  Bassett.
\newblock Stimulation-based control of dynamic brain networks.
\newblock \emph{PLoS Comput Biol}, 12\penalty0 (9):\penalty0 e1005076, 2016.

\bibitem[Betzel et~al.(2016)Betzel, Gu, Medaglia, Pasqualetti, and
  Bassett]{betzel2016optimally}
R~F Betzel, S~Gu, J~D Medaglia, F~Pasqualetti, and D~S Bassett.
\newblock Optimally controlling the human connectome: the role of network
  topology.
\newblock \emph{Sci Rep}, 6:\penalty0 30770, 2016.

\bibitem[Gu et~al.(2017)Gu, Betzel, Mattar, Cieslak, Delio, Grafton,
  Pasqualetti, and Bassett]{gu2017optimal}
S~Gu, R~F Betzel, M~G Mattar, M~Cieslak, P~R Delio, S~T Grafton, F~Pasqualetti,
  and D~S Bassett.
\newblock Optimal trajectories of brain state transitions.
\newblock \emph{Neuroimage}, 148:\penalty0 305--317, 2017.

\bibitem[Wiles et~al.(2017)Wiles, Gu, Pasqualetti, Bassett, and
  Meaney]{wiles2017autaptic}
L~Wiles, S~Gu, F~Pasqualetti, D~S Bassett, and D~F Meaney.
\newblock Autaptic connections shift network excitability and bursting.
\newblock \emph{Scientific Reports}, In Press, 2017.

\bibitem[Kim et~al.(2017)Kim, Soffer, Kahn, Vettel, Pasqualetti, and
  Bassett]{kim2017role}
J~Kim, J~M Soffer, A~E Kahn, J~M Vettel, F~Pasqualetti, and D~S Bassett.
\newblock Role of graph architecture in controlling dynamical networks with
  applications to neural systems.
\newblock \emph{Nature Physics}, Epub Ahead of Print, 2017.

\bibitem[Gu et~al.(2015)Gu, Pasqualetti, Cieslak, Telesford, Alfred, Kahn,
  Medaglia, Vettel, Miller, Grafton, et~al.]{gu2015controllability}
Shi Gu, Fabio Pasqualetti, Matthew Cieslak, Qawi~K Telesford, B~Yu Alfred,
  Ari~E Kahn, John~D Medaglia, Jean~M Vettel, Michael~B Miller, Scott~T
  Grafton, et~al.
\newblock Controllability of structural brain networks.
\newblock \emph{Nature communications}, 6, 2015.

\bibitem[Tang et~al.(2017)Tang, Giusti, Baum, Gu, Pollock, Kahn, Roalf, Moore,
  Ruparel, Gur, Gur, Satterthwaite, and Bassett]{tang2017developmental}
E~Tang, C~Giusti, G~L Baum, S~Gu, E~Pollock, A~E Kahn, D~R Roalf, T~M Moore,
  K~Ruparel, R~C Gur, R~E Gur, T~D Satterthwaite, and D~S Bassett.
\newblock Developmental increases in white matter network controllability
  support a growing diversity of brain dynamics.
\newblock \emph{Nat Commun}, 8\penalty0 (1):\penalty0 1252, 2017.

\bibitem[Satterthwaite et~al.(2014{\natexlab{b}})Satterthwaite, Elliott,
  Ruparel, Loughead, Prabhakaran, Calkins, Hopson, Jackson, Keefe, Riley,
  Mentch, Sleiman, Verma, Davatzikos, Hakonarson, Gur, and
  Gur]{satterthwaite2014neuroimaging}
T~D Satterthwaite, M~A Elliott, K~Ruparel, J~Loughead, K~Prabhakaran, M~E
  Calkins, R~Hopson, C~Jackson, J~Keefe, M~Riley, F~D Mentch, P~Sleiman,
  R~Verma, C~Davatzikos, H~Hakonarson, R~C Gur, and R~E Gur.
\newblock Neuroimaging of the philadelphia neurodevelopmental cohort.
\newblock \emph{Neuroimage}, 86:\penalty0 544--553, 2014{\natexlab{b}}.

\bibitem[Cammoun et~al.(2012)Cammoun, Gigandet, Meskaldji, Thiran, Sporns, Do,
  Maeder, Meuli, and Hagmann]{cammoun2012mapping}
L~Cammoun, X~Gigandet, D~Meskaldji, J~P Thiran, O~Sporns, K~Q Do, P~Maeder,
  R~Meuli, and P~Hagmann.
\newblock Mapping the human connectome at multiple scales with diffusion
  spectrum {MRI}.
\newblock \emph{J Neurosci Methods}, 203\penalty0 (2):\penalty0 386--397, 2012.

\bibitem[Roalf et~al.(2016)Roalf, Quarmley, Elliott, Satterthwaite, Vandekar,
  Ruparel, Gennatas, Calkins, Moore, Hopson, et~al.]{roalf2016impact}
David~R Roalf, Megan Quarmley, Mark~A Elliott, Theodore~D Satterthwaite,
  Simon~N Vandekar, Kosha Ruparel, Efstathios~D Gennatas, Monica~E Calkins,
  Tyler~M Moore, Ryan Hopson, et~al.
\newblock The impact of quality assurance assessment on diffusion tensor
  imaging outcomes in a large-scale population-based cohort.
\newblock \emph{Neuroimage}, 125:\penalty0 903--919, 2016.

\bibitem[Gur et~al.(2010)Gur, Richard, Hughett, Calkins, Macy, Bilker,
  Brensinger, and Gur]{gur2010cognitive}
R~C Gur, J~Richard, P~Hughett, M~E Calkins, L~Macy, W~B Bilker, C~Brensinger,
  and R~E Gur.
\newblock A cognitive neuroscience-based computerized battery for efficient
  measurement of individual differences: standardization and initial construct
  validation.
\newblock \emph{J Neurosci Methods}, 187\penalty0 (2):\penalty0 254--262, 2010.

\bibitem[Gur et~al.(2012)Gur, Richard, Calkins, Chiavacci, Hansen, Bilker,
  Loughead, Connolly, Qiu, Mentch, et~al.]{gur2012age}
Ruben~C Gur, Jan Richard, Monica~E Calkins, Rosetta Chiavacci, John~A Hansen,
  Warren~B Bilker, James Loughead, John~J Connolly, Haijun Qiu, Frank~D Mentch,
  et~al.
\newblock Age group and sex differences in performance on a computerized
  neurocognitive battery in children age 8- 21.
\newblock \emph{Neuropsychology}, 26\penalty0 (2):\penalty0 251, 2012.

\bibitem[Moore et~al.(2015)Moore, Reise, Gur, Hakonarson, and
  Gur]{moore2015psychometric}
Tyler~M Moore, Steven~P Reise, Raquel~E Gur, Hakon Hakonarson, and Ruben~C Gur.
\newblock Psychometric properties of the penn computerized neurocognitive
  battery.
\newblock \emph{Neuropsychology}, 29\penalty0 (2):\penalty0 235, 2015.

\bibitem[Moore et~al.(2016)Moore, Reise, Roalf, Satterthwaite, Davatzikos,
  Bilker, Port, Jackson, Ruparel, Savitt, et~al.]{moore2016development}
Tyler~M Moore, Steven~P Reise, David~R Roalf, Theodore~D Satterthwaite,
  Christos Davatzikos, Warren~B Bilker, Allison~M Port, Chad~T Jackson, Kosha
  Ruparel, Adam~P Savitt, et~al.
\newblock Development of an itemwise efficiency scoring method: Concurrent,
  convergent, discriminant, and neuroimaging-based predictive validity assessed
  in a large community sample.
\newblock \emph{Psychol Assess.}, 28\penalty0 (12):\penalty0 1529--1542, 2016.

\bibitem[Moore et~al.(2017)Moore, Gur, Thomas, Brown, Nock, Savitt, Keilp,
  Heeringa, Ursano, Stein, et~al.]{moore2017development}
Tyler~M Moore, Ruben~C Gur, Michael~L Thomas, Gregory~G Brown, Matthew~K Nock,
  Adam~P Savitt, John~G Keilp, Steven Heeringa, Robert~J Ursano, Murray~B
  Stein, et~al.
\newblock Development, administration, and structural validity of a brief,
  computerized neurocognitive battery: Results from the army study to assess
  risk and resilience in service members.
\newblock \emph{Assessment}, Epub Ahead of Print, 2017.

\bibitem[Vandekar et~al.(2015)Vandekar, Shinohara, Raznahan, Roalf, Ross,
  DeLeo, Ruparel, Verma, Wolf, Gur, et~al.]{vandekar2015topologically}
Simon~N Vandekar, Russell~T Shinohara, Armin Raznahan, David~R Roalf, Michelle
  Ross, Nicholas DeLeo, Kosha Ruparel, Ragini Verma, Daniel~H Wolf, Ruben~C
  Gur, et~al.
\newblock Topologically dissociable patterns of development of the human
  cerebral cortex.
\newblock \emph{Journal of Neuroscience}, 35\penalty0 (2):\penalty0 599--609,
  2015.

\bibitem[Fischl(2012)]{fischl2012freesurfer}
B~Fischl.
\newblock Freesurfer.
\newblock \emph{Neuroimage}, 62:\penalty0 774--781, 2012.

\bibitem[Rosen et~al.(2017)Rosen, Roalf, Ruparel, Blake, Seelaus, Villa, Ciric,
  Cook, Davatzikos, Elliott, de~La~Garza, Gennatas, Quarmley, Schmitt,
  Shinohara, Tisdall, Craddock, Gur, Gur, and Satterthwaite]{Rosen2017}
Adon~F.G. Rosen, David~R. Roalf, Kosha Ruparel, Jason Blake, Kevin Seelaus,
  Lakshmi~P. Villa, Rastko Ciric, Philip~A. Cook, Christos Davatzikos, Mark~A.
  Elliott, Angel~Garcia de~La~Garza, Efstathios~D. Gennatas, Megan Quarmley,
  J.~Eric Schmitt, Russell~T. Shinohara, M.~Dylan Tisdall, R.~Cameron Craddock,
  Raquel~E. Gur, Ruben~C. Gur, and Theodore~D. Satterthwaite.
\newblock Data-driven assessment of structural image quality.
\newblock \emph{NeuroImage}, pages~--, 2017.
\newblock ISSN 1053-8119.
\newblock \doi{https://doi.org/10.1016/j.neuroimage.2017.12.059}.
\newblock URL
  \url{https://www.sciencedirect.com/science/article/pii/S1053811917310832}.

\bibitem[Cuadra et~al.(2004)Cuadra, Pollo, Bardera, Cuisenaire, Villemure, and
  Thiran]{lausanne}
Meritxell~Bach Cuadra, Claudio Pollo, Anton Bardera, Olivier Cuisenaire, J-G
  Villemure, and J-P Thiran.
\newblock Atlas-based segmentation of pathological mr brain images using a
  model of lesion growth.
\newblock \emph{IEEE transactions on medical imaging}, 23\penalty0
  (10):\penalty0 1301--1314, 2004.

\bibitem[Ragland et~al.(2002)Ragland, Turetsky, Gur, Gunning-Dixon, Turner,
  Schroeder, Chan, and Gur]{ragland2002working}
J~Daniel Ragland, Bruce~I Turetsky, Ruben~C Gur, Faith Gunning-Dixon, Travis
  Turner, Lee Schroeder, Robin Chan, and Raquel~E Gur.
\newblock Working memory for complex figures: an fmri comparison of letter and
  fractal n-back tasks.
\newblock \emph{Neuropsychology}, 16\penalty0 (3):\penalty0 370, 2002.

\bibitem[Shanmugan et~al.(2016)Shanmugan, Wolf, Calkins, Moore, Ruparel,
  Hopson, Vandekar, Roalf, Elliott, Jackson, et~al.]{shanmugan2016common}
Sheila Shanmugan, Daniel~H Wolf, Monica~E Calkins, Tyler~M Moore, Kosha
  Ruparel, Ryan~D Hopson, Simon~N Vandekar, David~R Roalf, Mark~A Elliott, Chad
  Jackson, et~al.
\newblock Common and dissociable mechanisms of executive system dysfunction
  across psychiatric disorders in youth.
\newblock \emph{American journal of psychiatry}, 173\penalty0 (5):\penalty0
  517--526, 2016.

\bibitem[Jones and Basser(2004)]{jones2004squashing}
D~K Jones and P~J Basser.
\newblock ``squashing peanuts and smashing pumpkins'': how noise distorts
  diffusion-weighted {MR} data.
\newblock \emph{Magn. Reson. Med.}, 52:\penalty0 979--993, 2004.

\bibitem[Chen et~al.(2015)Chen, Tymofiyeva, Hess, and Xu]{chen2015effects}
Y~Chen, O~Tymofiyeva, C~P Hess, and D~Xu.
\newblock Effects of rejecting diffusion directions on tensor-derived
  parameters.
\newblock \emph{Neuroimage}, 109:\penalty0 160--170, 2015.

\bibitem[Jenkinson et~al.(2002)Jenkinson, Bannister, Brady, and
  Smith]{jenkinson2002improved}
M~Jenkinson, P~Bannister, M~Brady, and S~Smith.
\newblock Improved optimization for the robust and accurate linear registration
  and motion correction of brain images.
\newblock \emph{Neuroimage}, 17:\penalty0 825--841, 2002.

\bibitem[Andersson and Sotiropoulos(2016)]{andersson2016integrated}
J~L~R Andersson and S~N Sotiropoulos.
\newblock An integrated approach to correction for off-resonance effects and
  subject movement in diffusion {MR} imaging.
\newblock \emph{Neuroimage}, 125:\penalty0 1063--1078, 2016.

\bibitem[Jenkinson et~al.(2012)Jenkinson, Beckmann, Behrens, Woolrich, and
  Smith]{jenkinson2012fsl}
M~Jenkinson, C~F Beckmann, T~E Behrens, M~W Woolrich, and S~M Smith.
\newblock Fsl.
\newblock \emph{Neuroimage}, 62:\penalty0 782--790, 2012.

\bibitem[Satterthwaite et~al.(2013)Satterthwaite, Wolf, Erus, Ruparel, Elliott,
  Gennatas, Hopson, Jackson, Prabhakaran, Bilker,
  et~al.]{satterthwaite2013functional}
Theodore~D Satterthwaite, Daniel~H Wolf, Guray Erus, Kosha Ruparel, Mark~A
  Elliott, Efstathios~D Gennatas, Ryan Hopson, Chad Jackson, Karthik
  Prabhakaran, Warren~B Bilker, et~al.
\newblock Functional maturation of the executive system during adolescence.
\newblock \emph{Journal of Neuroscience}, 33\penalty0 (41):\penalty0
  16249--16261, 2013.

\bibitem[Greve and Fischl(2009)]{greve2009accurate}
Douglas~N Greve and Bruce Fischl.
\newblock Accurate and robust brain image alignment using boundary-based
  registration.
\newblock \emph{Neuroimage}, 48\penalty0 (1):\penalty0 63--72, 2009.

\bibitem[Honey et~al.(2009)Honey, Sporns, Cammoun, Gigandet, Thiran, Meuli, and
  Hagmann]{honey2009predicting}
CJ~Honey, O~Sporns, Leila Cammoun, Xavier Gigandet, Jean-Philippe Thiran, Reto
  Meuli, and Patric Hagmann.
\newblock Predicting human resting-state functional connectivity from
  structural connectivity.
\newblock \emph{Proceedings of the National Academy of Sciences}, 106\penalty0
  (6):\penalty0 2035--2040, 2009.

\bibitem[Pecora and Carroll(1998)]{pecora1998master}
Louis~M Pecora and Thomas~L Carroll.
\newblock Master stability functions for synchronized coupled systems.
\newblock \emph{Physical review letters}, 80\penalty0 (10):\penalty0 2109,
  1998.

\bibitem[Benjamini and Hochberg(1995)]{bhfdr}
Yoav Benjamini and Yosef Hochberg.
\newblock Controlling the false discovery rate: a practical and powerful
  approach to multiple testing.
\newblock \emph{Journal of the royal statistical society. Series B
  (Methodological)}, pages 289--300, 1995.

\bibitem[Wood(2004)]{wood2004stable}
Simon~N Wood.
\newblock Stable and efficient multiple smoothing parameter estimation for
  generalized additive models.
\newblock \emph{Journal of the American Statistical Association}, 99\penalty0
  (467):\penalty0 673--686, 2004.

\bibitem[Tingley et~al.(2014)Tingley, Yamamoto, Hirose, Keele, and
  Imai]{tingley2014mediation}
Dustin Tingley, Teppei Yamamoto, Kentaro Hirose, Luke Keele, and Kosuke Imai.
\newblock Mediation: R package for causal mediation analysis.
\newblock 2014.

\bibitem[Preacher et~al.(2004)Preacher, Curran, and Bauer]{preacher2004simple}
KJ~Preacher, PJ~Curran, and DJ~Bauer.
\newblock Simple intercepts, simple slopes, and regions of significance in mlr
  2-way interactions. retrieved september 25, 2005, 2004.

\bibitem[Snodgrass and Corwin(1988)]{snodgrass1988pragmatics}
Joan~G Snodgrass and June Corwin.
\newblock Pragmatics of measuring recognition memory: applications to dementia
  and amnesia.
\newblock \emph{Journal of Experimental Psychology: General}, 117\penalty0
  (1):\penalty0 34, 1988.

\bibitem[Castellanos and Tannock(2002)]{castellanos2002neuroscience}
F~Xavier Castellanos and Rosemary Tannock.
\newblock Neuroscience of attention-deficit/hyperactivity disorder: the search
  for endophenotypes.
\newblock \emph{Nature Reviews Neuroscience}, 3\penalty0 (8):\penalty0
  617--628, 2002.

\bibitem[Pohjalainen et~al.(1998)Pohjalainen, Rinne, N{\aa}gren, Syv{\"a}lahti,
  and Hietala]{pohjalainen1998sex}
Tiina Pohjalainen, Juha~O Rinne, Kjell N{\aa}gren, Erkka Syv{\"a}lahti, and
  Jarmo Hietala.
\newblock Sex differences in the striatal dopamine d2 receptor binding
  characteristics in vivo.
\newblock \emph{American Journal of Psychiatry}, 155\penalty0 (6):\penalty0
  768--773, 1998.

\bibitem[Stephan et~al.(2017)Stephan, Schlagenhauf, Huys, Raman, Aponte,
  Brodersen, Rigoux, Moran, Daunizeau, Dolan, Friston, and
  Heinz]{STEPHAN2017180}
K.E. Stephan, F.~Schlagenhauf, Q.J.M. Huys, S.~Raman, E.A. Aponte, K.H.
  Brodersen, L.~Rigoux, R.J. Moran, J.~Daunizeau, R.J. Dolan, K.J. Friston, and
  A.~Heinz.
\newblock Computational neuroimaging strategies for single patient predictions.
\newblock \emph{NeuroImage}, 145\penalty0 (Part B):\penalty0 180 -- 199, 2017.
\newblock ISSN 1053-8119.
\newblock \doi{https://doi.org/10.1016/j.neuroimage.2016.06.038}.
\newblock URL
  \url{http://www.sciencedirect.com/science/article/pii/S1053811916302877}.
\newblock Individual Subject Prediction.

\bibitem[Izhikevich and Edelman(2008)]{izhikevich2008large}
Eugene~M Izhikevich and Gerald~M Edelman.
\newblock Large-scale model of mammalian thalamocortical systems.
\newblock \emph{Proceedings of the national academy of sciences}, 105\penalty0
  (9):\penalty0 3593--3598, 2008.

\bibitem[Murphy and Bassett(2017)]{murphy2017network}
A~C Murphy and D~S Bassett.
\newblock A network neuroscience of neurofeedback for clinical translation.
\newblock \emph{Curr Opin Biomed Eng}, 1:\penalty0 63--70, 2017.

\bibitem[Tanji and Hoshi(2008)]{tanji2008role}
Jun Tanji and Eiji Hoshi.
\newblock Role of the lateral prefrontal cortex in executive behavioral
  control.
\newblock \emph{Physiological reviews}, 88\penalty0 (1):\penalty0 37--57, 2008.

\bibitem[Euston et~al.(2012)Euston, Gruber, and McNaughton]{euston2012role}
David~R Euston, Aaron~J Gruber, and Bruce~L McNaughton.
\newblock The role of medial prefrontal cortex in memory and decision making.
\newblock \emph{Neuron}, 76\penalty0 (6):\penalty0 1057--1070, 2012.

\bibitem[JL(1993)]{cummings1993frontsubc}
Cummings JL.
\newblock Frontal-subcortical circuits and human behavior.
\newblock \emph{Archives of Neurology}, 50\penalty0 (8):\penalty0 873--880,
  1993.
\newblock \doi{10.1001/archneur.1993.00540080076020}.
\newblock URL \url{+ http://dx.doi.org/10.1001/archneur.1993.00540080076020}.

\bibitem[Duke and Kaszniak(2000)]{duke2000executive}
Lisa~M Duke and Alfred~W Kaszniak.
\newblock Executive control functions in degenerative dementias: A comparative
  review.
\newblock \emph{Neuropsychology review}, 10\penalty0 (2):\penalty0 75--99,
  2000.

\bibitem[Bonelli and Cummings(2007)]{bonelli2007frontal}
Raphael~M Bonelli and Jeffrey~L Cummings.
\newblock Frontal-subcortical circuitry and behavior.
\newblock \emph{Dialogues in clinical neuroscience}, 9\penalty0 (2):\penalty0
  141, 2007.

\bibitem[Tekin and Cummings(2002)]{tekin2002frontal}
Sibel Tekin and Jeffrey~L Cummings.
\newblock Frontal--subcortical neuronal circuits and clinical neuropsychiatry:
  an update.
\newblock \emph{Journal of psychosomatic research}, 53\penalty0 (2):\penalty0
  647--654, 2002.

\bibitem[Bombois et~al.(2007)Bombois, Debette, Delbeuck, Bruandet, Lepoittevin,
  Delmaire, Leys, and Pasquier]{Bombois2595}
St{\'e}phanie Bombois, St{\'e}phanie Debette, Xavier Delbeuck, Am{\'e}lie
  Bruandet, Samuel Lepoittevin, Christine Delmaire, Didier Leys, and Florence
  Pasquier.
\newblock Prevalence of subcortical vascular lesions and association with
  executive function in mild cognitive impairment subtypes.
\newblock \emph{Stroke}, 38\penalty0 (9):\penalty0 2595--2597, 2007.
\newblock ISSN 0039-2499.
\newblock \doi{10.1161/STROKEAHA.107.486407}.
\newblock URL \url{http://stroke.ahajournals.org/content/38/9/2595}.

\bibitem[Plichta et~al.(2009)Plichta, Vasic, Wolf, Lesch, Brummer, Jacob,
  Fallgatter, and Grön]{PLICHTA20097}
Michael~M. Plichta, Nenad Vasic, Robert~Christian Wolf, Klaus-Peter Lesch,
  Dagmar Brummer, Christian Jacob, Andreas~J. Fallgatter, and Georg Grön.
\newblock Neural hyporesponsiveness and hyperresponsiveness during immediate
  and delayed reward processing in adult attention-deficit/hyperactivity
  disorder.
\newblock \emph{Biological Psychiatry}, 65\penalty0 (1):\penalty0 7 -- 14,
  2009.
\newblock ISSN 0006-3223.
\newblock \doi{https://doi.org/10.1016/j.biopsych.2008.07.008}.
\newblock URL
  \url{http://www.sciencedirect.com/science/article/pii/S0006322308008275}.
\newblock Perception, Empathy, and Reward in Attention-Deficit/Hyperactivity
  Disorder and Autism.

\bibitem[Cubillo et~al.(2012)Cubillo, Halari, Smith, Taylor, and
  Rubia]{CUBILLO2012194}
Ana Cubillo, Rozmin Halari, Anna Smith, Eric Taylor, and Katya Rubia.
\newblock A review of fronto-striatal and fronto-cortical brain abnormalities
  in children and adults with attention deficit hyperactivity disorder (adhd)
  and new evidence for dysfunction in adults with adhd during motivation and
  attention.
\newblock \emph{Cortex}, 48\penalty0 (2):\penalty0 194 -- 215, 2012.
\newblock ISSN 0010-9452.
\newblock \doi{https://doi.org/10.1016/j.cortex.2011.04.007}.
\newblock URL
  \url{http://www.sciencedirect.com/science/article/pii/S001094521100102X}.
\newblock Frontal lobes.

\bibitem[Vaidya(2011)]{vaidya2011neurodevelopmental}
Chandan~J Vaidya.
\newblock Neurodevelopmental abnormalities in adhd.
\newblock In \emph{Behavioral neuroscience of attention deficit hyperactivity
  disorder and its treatment}, pages 49--66. Springer, 2011.

\bibitem[Raz et~al.(1995)Raz, Torres, and Acker]{raz1995age}
Naftali Raz, Ivan~J Torres, and James~D Acker.
\newblock Age, gender, and hemispheric differences in human striatum: a
  quantitative review and new data from in vivo mri morphometry.
\newblock \emph{Neurobiology of learning and memory}, 63\penalty0 (2):\penalty0
  133--142, 1995.

\bibitem[Freeman and Skarda(1985)]{FREEMAN1985147}
Walter~J. Freeman and Christine~A. Skarda.
\newblock Spatial eeg patterns, non-linear dynamics and perception: the
  neo-sherringtonian view.
\newblock \emph{Brain Research Reviews}, 10\penalty0 (3):\penalty0 147 -- 175,
  1985.
\newblock ISSN 0165-0173.
\newblock \doi{https://doi.org/10.1016/0165-0173(85)90022-0}.
\newblock URL
  \url{http://www.sciencedirect.com/science/article/pii/0165017385900220}.

\bibitem[Zhu et~al.(2011)Zhu, Hu, Qiu, Taylor, Tso, Yiannoutsos, Navia, Mori,
  Ekholm, Schifitto, et~al.]{zhu2011quantification}
Tong Zhu, Rui Hu, Xing Qiu, Michael Taylor, Yuen Tso, Constantin Yiannoutsos,
  Bradford Navia, Susumu Mori, Sven Ekholm, Giovanni Schifitto, et~al.
\newblock Quantification of accuracy and precision of multi-center dti
  measurements: a diffusion phantom and human brain study.
\newblock \emph{Neuroimage}, 56\penalty0 (3):\penalty0 1398--1411, 2011.

\bibitem[Maier-Hein et~al.(2017)Maier-Hein, Neher, Houde, C{\^o}t{\'e},
  Garyfallidis, Zhong, Chamberland, Yeh, Lin, Ji, et~al.]{maier2017challenge}
Klaus~H Maier-Hein, Peter~F Neher, Jean-Christophe Houde, Marc-Alexandre
  C{\^o}t{\'e}, Eleftherios Garyfallidis, Jidan Zhong, Maxime Chamberland,
  Fang-Cheng Yeh, Ying-Chia Lin, Qing Ji, et~al.
\newblock The challenge of mapping the human connectome based on diffusion
  tractography.
\newblock \emph{Nature communications}, 8:\penalty0 1349, 2017.

\bibitem[Kubicki et~al.(2007)Kubicki, McCarley, Westin, Park, Maier, Kikinis,
  Jolesz, and Shenton]{kubicki2007review}
Marek Kubicki, Robert McCarley, Carl-Fredrik Westin, Hae-Jeong Park, Stephan
  Maier, Ron Kikinis, Ferenc~A Jolesz, and Martha~E Shenton.
\newblock A review of diffusion tensor imaging studies in schizophrenia.
\newblock \emph{Journal of psychiatric research}, 41\penalty0 (1):\penalty0
  15--30, 2007.

\bibitem[Pievani et~al.(2011)Pievani, de~Haan, Wu, Seeley, and
  Frisoni]{pievani2011functional}
Michela Pievani, Willem de~Haan, Tao Wu, William~W Seeley, and Giovanni~B
  Frisoni.
\newblock Functional network disruption in the degenerative dementias.
\newblock \emph{The Lancet Neurology}, 10\penalty0 (9):\penalty0 829--843,
  2011.

\bibitem[Joel et~al.(2015)Joel, Berman, Tavor, Wexler, Gaber, Stein, Shefi,
  Pool, Urchs, Margulies, et~al.]{joel2015sex}
Daphna Joel, Zohar Berman, Ido Tavor, Nadav Wexler, Olga Gaber, Yaniv Stein,
  Nisan Shefi, Jared Pool, Sebastian Urchs, Daniel~S Margulies, et~al.
\newblock Sex beyond the genitalia: The human brain mosaic.
\newblock \emph{Proceedings of the National Academy of Sciences}, 112\penalty0
  (50):\penalty0 15468--15473, 2015.

\bibitem[Joel and McCarthy(2017)]{joel2017incorporating}
Daphna Joel and Margaret~M McCarthy.
\newblock Incorporating sex as a biological variable in neuropsychiatric
  research: where are we now and where should we be?
\newblock \emph{Neuropsychopharmacology}, 42\penalty0 (2):\penalty0 379--385,
  2017.

\end{thebibliography}

\newpage
\section*{Figures}
\selectlanguage{english}
\selectlanguage{english}

\newpage
\selectlanguage{english}

\newpage
\selectlanguage{english}

\newpage
\selectlanguage{english}

\newpage
\selectlanguage{english}

\newpage
\selectlanguage{english}
\newpage

\newpage


\selectlanguage{english}
\clearpage

\newpage
\appendix
\counterwithin{figure}{section}
\section{Supplementary Data}

\newpage
\newpage

\newpage

\newpage

\newpage
\begin{figure}[h!]
	\centering
	\begin{minipage}{\linewidth}
		\begin{center}
			\title{a. \textit{Weighted Degree}}
		\end{center}
		\begin{tabular}{rrrrrrrrr}
			\hline
			& ACME & p & ADE & p & Total Effect & p & Prop. Mediated & p \\ 
			\hline
			R Medial orbito frontal$_1$ & 0.01 & 0.54 & 0.15 & 0.00 & 0.16 & 0.00 & 0.06 & 0.55 \\ 
			R Rostral middle frontal$_6$ & 0.01 & 0.25 & 0.15 & 0.00 & 0.16 & 0.00 & 0.06 & 0.27 \\ 
			R Superior frontal$_4$ & 0.00 & 0.79 & 0.15 & 0.00 & 0.15 & 0.00 & 0.02 & 0.79 \\ 
			R Postcentral$_5$ & 0.01 & 0.77 & 0.15 & 0.00 & 0.15 & 0.00 & 0.05 & 0.77 \\ 
			R Superior parietal$_1$ & -0.00 & 0.91 & 0.16 & 0.00 & 0.15 & 0.00 & -0.00 & 0.91 \\ 
			R Inferior parietal$_1$ & -0.00 & 0.91 & 0.15 & 0.00 & 0.15 & 0.00 & -0.00 & 0.91 \\ 
			R Inferior parietal$_3$ & -0.00 & 0.91 & 0.16 & 0.00 & 0.15 & 0.00 & -0.01 & 0.91 \\ 
			R Lateral occipital$_1$ & -0.01 & 0.54 & 0.16 & 0.00 & 0.16 & 0.01 & -0.04 & 0.55 \\ 
			R Fusiform$_4$ & 0.00 & 0.85 & 0.15 & 0.00 & 0.16 & 0.00 & 0.03 & 0.86 \\ 
			R Caudate & 0.01 & 0.06 & 0.15 & 0.00 & 0.16 & 0.00 & 0.08 & 0.12 \\ 
			L Lateral orbitofrontal$_3$ & -0.01 & 0.25 & 0.17 & 0.00 & 0.16 & 0.00 & -0.07 & 0.27 \\ 
			L Pars triangularis$_1$ & -0.00 & 0.91 & 0.16 & 0.00 & 0.16 & 0.00 & -0.00 & 0.91 \\ 
			L Precentral$_4$ & 0.00 & 0.91 & 0.16 & 0.00 & 0.16 & 0.00 & 0.01 & 0.91 \\ 
			L Precentral$_8$ & 0.00 & 0.91 & 0.16 & 0.00 & 0.16 & 0.00 & 0.01 & 0.91 \\ 
			L Postcentral$_1$ & -0.00 & 0.91 & 0.16 & 0.00 & 0.16 & 0.00 & -0.01 & 0.91 \\ 
			L Postcentral$_3$ & 0.00 & 0.91 & 0.15 & 0.00 & 0.16 & 0.00 & 0.00 & 0.91 \\ 
			L Postcentral$_4$ & -0.00 & 0.91 & 0.16 & 0.00 & 0.16 & 0.00 & -0.01 & 0.91 \\ 
			L Superior parietal$_2$ & -0.01 & 0.54 & 0.17 & 0.00 & 0.16 & 0.00 & -0.03 & 0.55 \\ 
			L Superior parietal$_5$ & -0.00 & 0.91 & 0.16 & 0.00 & 0.15 & 0.00 & -0.01 & 0.91 \\ 
			L Inferior parietal$_3$ & 0.00 & 0.86 & 0.16 & 0.00 & 0.16 & 0.00 & 0.01 & 0.86 \\ 
			L Lateral occipital$_4$ & 0.00 & 0.91 & 0.16 & 0.00 & 0.16 & 0.00 & 0.00 & 0.91 \\ 
			L Lateral occipital$_5$ & 0.00 & 0.86 & 0.15 & 0.00 & 0.16 & 0.00 & 0.02 & 0.86 \\ 
			L Fusiform$_4$ & 0.01 & 0.78 & 0.15 & 0.00 & 0.16 & 0.00 & 0.05 & 0.78 \\ 
			L Superior temporal$_4$ & -0.01 & 0.70 & 0.16 & 0.00 & 0.16 & 0.00 & -0.04 & 0.70 \\ 
			L Insula$_3$ & 0.00 & 0.86 & 0.15 & 0.00 & 0.16 & 0.00 & 0.01 & 0.86 \\ 
			L Caudate & 0.01 & 0.54 & 0.15 & 0.00 & 0.16 & 0.00 & 0.06 & 0.55 \\ 
			L Putamen & 0.00 & 0.70 & 0.15 & 0.00 & 0.16 & 0.00 & 0.02 & 0.70 \\ 
			L Pallidum & 0.01 & 0.61 & 0.15 & 0.00 & 0.16 & 0.00 & 0.04 & 0.61 \\ 
			L Amygdala & 0.01 & 0.54 & 0.15 & 0.00 & 0.16 & 0.00 & 0.04 & 0.55 \\ 
			\hline
		\end{tabular}
	\end{minipage}
\bigskip
\caption{{\textbf{Sex-(weighted degree)-executive function mediation relationship.} \emph{(a-b)} Non-parametric causal mediation analysis for sex $\rightarrow$ weighted degree $\rightarrow$ executive function, using control nodes with sex-associated average controllability, modal controllability, or weighted degree. ACME = average causal mediation effect, ADE = average direct effect. We performed 1000 simulations. The $p$-values were FDR corrected ($q < 0.05$). Age, total brain volume, handedness, and head motion were included in all analyses.
		\label{fig:figureS6}%
}}
\end{figure}

\end{document}